\documentclass[reprint,amsmath,amsfonts,amssymb,aps,prl,longbibliography]{revtex4-1}
\usepackage{graphicx}
\usepackage{siunitx}
\usepackage{braket}

\begin{document}

\title{Deterministic generation of a two-dimensional cluster state}
\author{Mikkel V. Larsen}
\email{mivila@fysik.dtu.dk}
\author{Xueshi Guo}
\author{Casper R. Breum}
\author{Jonas S. Neergaard-Nielsen}
\author{Ulrik L. Andersen}
\email{ulrik.andersen@fysik.dtu.dk}
\affiliation{Center for Macroscopic Quantum States (bigQ), Department of Physics, Technical University of Denmark, Fysikvej, 2800 Kgs. Lyngby, Denmark}
\date{September 20, 2019}

\begin{abstract}
Measurement-based quantum computation offers exponential computational speed-up via simple measurements on a large entangled cluster state. We propose and demonstrate a scalable scheme for the generation of photonic cluster states suitable for universal measurement-based quantum computation. We exploit temporal multiplexing of squeezed light modes, delay loops, and beam-splitter transformations to deterministically generate a cylindrical cluster state with a two-dimensional (2D) topological structure as required for universal quantum information processing. The generated state consists of more than $\SI{30000}{}$ entangled modes arranged in a cylindrical lattice with 24 modes on the circumference, defining the input register, and a length of $\SI{1250}{}$ modes, defining the computation depth. Our demonstrated source of 2D cluster states can be combined with quantum error correction to enable fault-tolerant quantum computation.
\end{abstract}

\maketitle

%%% Body %%%
Quantum computing represents a new paradigm for information processing that harnesses the inherent non-classical features of quantum physics to find solutions to problems that are computationally intractable on classical processors \cite{ladd10}. In measurement-based, or cluster state, quantum computing (MBQC), the processing is performed via simple single-site measurements on a large entangled cluster state \cite{raussendorf01}. This constitutes a simplification over the standard gate-based model of quantum computing, as it replaces complex coherent unitary dynamics with simple projective measurements. However, one of the outstanding challenges in realizing cluster state computation is the reliable, deterministic and scalable generation of non-classical entangled states suitable for universal information processing. 

Several candidate platforms for scalable cluster state generation have been proposed and some experimentally realized, including solid state superconducting qubits \cite{wang18}, trapped ion qubits \cite{mandel03,lanyon13} and photonic qubits or qumodes, in which qubits can be encoded, generated by parametric down-conversion \cite{walther05,tokunaga08,yokoyama13,chen14} or by quantum dots \cite{schwartz16}. However, none of these implementations have demonstrated true scalability combined with computational universality. The largest cluster state generated to date is a temporally multiplexed photonic state comprising entangled modes in a long chain which however does not allow for universal computation due to its one-dimensional (1D) topological structure \cite{yokoyama13,yoshikawa16}. To achieve universality, the dimension of the cluster state must be at least two. Several proposals for generating two-dimensional (2D) cluster states in different systems have been proposed \cite{economou10,menicucci11,wang14,alexander16} but due to technical challenges, scalable and computationally universal cluster states have yet to be produced in any physical system.

We propose and demonstrate a highly scalable scheme for the generation of cluster states for universal quantum computation based on quantum continuous variables (CV) where information is encoded in the position or momentum quadratures of photonic harmonic oscillators \cite{weedbrook12}. We use a temporally multiplexed source of optical Einstein-Podolsky-Rosen (EPR) states \cite{reid09} to generate a long string of entangled modes that is curled up and fused to form a 2D cylindrical array of entangled modes. Specifically, we generate a massive cluster state of more than $\SI{30000}{}$ entangled modes comprising an input register of $2\times12=24$ modes on which the input state may be encoded, and a length of $\SI{1250}{}$ modes for encoding operations by projective measurements, only limited by the phase stability of our setup. In addition to being universal and deterministically generated, the source is operated under ambient conditions in optical fibers at the low-loss telecom wavelength of $\SI{1550}{nm}$. These favorable operational conditions and specifications significantly facilitate further upscaling of the entangled state as well as its use in applications and fundamental studies. 

\begin{figure*}
	\includegraphics[width=\linewidth]{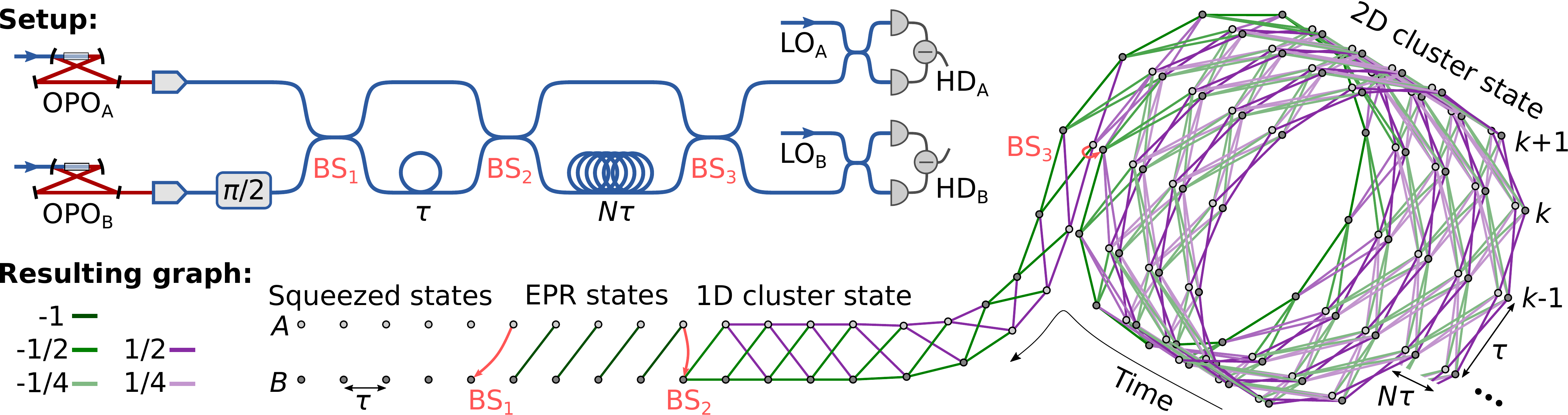}
	\caption{ Scheme of 2D cluster state generation. Squeezing is produced by two OPOs ($\text{OPO}_\text{A}$ and $\text{OPO}_\text{B}$), and coupled into fiber with 97\% coupling efficiency. There, temporal modes are interfered with fiber coupled beam splitters to generate a 2D cluster state. The corresponding graph is shown: Temporal modes of squeezing with mode index $k$ in two spatial modes $A$ and $B$ (bright and dark nodes) are interfered to generate EPR-states at $\text{BS}_1$. The EPR pairs are entangled to form a 1D cluster state using a $\tau$ delay in mode $B$ and $\text{BS}_2$, and the 1D cluster state is curled up to a 2D cluster state by another delay of $N\tau$ and $\text{BS}_3$.  Using homodyne detectors ($\text{HD}_\text{A}$ and $\text{HD}_\text{B}$), the temporal mode quadratures are measured from which the nullifiers are calculated. In the experimental implementation, the short delay is a $\SI{50.5}{m}$ fiber leading to temporal modes of $\SI{247}{ns}$ duration, while the long delay is a $\SI{606}{m}$ fiber such that $N=12$ as the illustrated graph. The temporal modes are defined by an asymmetric shaped temporal mode function within the $\SI{247}{ns}$ duration which filters out low frequency noise and leads to less than $10^{-3}$ mode overlap \cite{yoshikawa16}. For more information, see Material and Methods \cite{sm}.}
\end{figure*}

The canonical approach to CV cluster state generation is to apply two-mode controlled-Z gates onto pairs of individually prepared eigenstates of the momentum (or phase quadrature) operators $\hat{p}_i, \hat{p}_j$ in adjacent modes $i,j$. The gate is described by the unitary operation $\hat{C}_Z = e^{ig\hat{x}_i\hat{x}_j}$ where $\hat{x}_i,\hat{x}_j$ are the position (amplitude quadrature) operators of mode $i$ and $j$, while $g$ is the interaction strength. Applying this gate to two modes leads to entanglement in the form of quantum correlations of the two modes' quadratures. The operations and resulting state can be represented by a graph in which the nodes represent the momentum eigenstates while the edges (links) between the nodes represent the application of a controlled-Z operation where the interaction strength is given by the edge weight. In a practical implementation, the unphysical momentum eigenstates are replaced by highly squeezed states while the controlled-Z operations can be imitated by phase shifts and beam splitter transformations. To enable scalability, it has been suggested to use multiplexing of spatial modes \cite{armstrong12}, frequency modes \cite{menicucci08,cai17}, or temporal modes \cite{menicucci11,alexander18}. For example, Menicucci suggested using temporal multiplexing to form a 2D cluster state combining four squeezed state generators, five beam splitters, and two delay lines \cite{menicucci11}. 

We propose a simpler approach to 2D cluster state generation lowering the experimental requirements (Fig. 1). The state is produced in four steps: i) Pairs of squeezed vacuum states are generated at $\SI{1550}{nm}$ wavelength from two bow-tie shaped optical parametric oscillators (OPOs) by parametric down conversion \cite{andersen16}. The states are defined in consecutive temporal modes of duration $\tau$ of the continuously generated OPO output. ii) The squeezed vacuum pairs in spatial modes $A$ and $B$ are interfered on a balanced beam splitter (denoted $\text{BS}_1$). This produces a train of pairwise EPR-entangled temporal modes exhibiting quantum correlation between the position and momentum quadratures. Each EPR pair can be represented by a simple graph of a single edge connecting two nodes. iii) A 1D cluster state is formed by delaying one arm of the interferometer by $\tau$ with respect to the other arm and interfering the resulting time-synchronized modes on another balanced beam splitter (denoted $\text{BS}_2$). The interference entangles EPR pairs along an indefinitely long chain creating a 1D graph. iv) In the final step, the 2D cluster state is produced by introducing another delay to one interferometer arm of duration $N\tau$ and interfering the resulting time-synchronized modes on a final beam splitter (denoted $\text{BS}_3$). This effectively curls up the graph and fuses the modes into an indefinitely long cylinder with $N$ nodes on the circumference as illustrated in Fig.~1 for $N=12$, leading to $2\times N=24$ input modes distributed on the two spatial modes $A$ and $B$. For detailed description of experimental implementations see Material and Methods \cite{sm}.

All states and operations involved are Gaussian, meaning they can be described by Gaussian distributions of the quadrature variables in phase space. In the formalism of graphical calculus for Gaussian states \cite{menicucci11b}, the generated graphs are so-called $\mathcal{H}$-graphs as they can be generated from vacuum by a single Hamiltonian, and have an edge weight of $g=i\sinh(2r)G$ where $r$ is the squeezing parameter of the two squeezing operations and $G=-1$ for the EPR-states, $\pm1/2$ for the 1D graph and $\pm1/4,1/2$ for the 2D graph. Due to the particular structure of the $\mathcal{H}$-graph generated here (it is self-inverse and bipartite---see Supplementary Text section 1.1 for details \cite{sm}), it can be transformed into a cluster state by $\pi/2$ rotations in phase space leading to real edges of weight $g=\tanh(2r)G\rightarrow G$ for $r\rightarrow\infty$. Finally, as the $\pi/2$ phase space rotations can be absorbed into the measurement basis, or simply by appropriate re-definitions of quadratures on the rotated modes, the generated $\mathcal{H}$-graph state and its corresponding cluster state are completely equivalent. See Supplementary Text section 1.2 for details on the cluster state generation scheme \cite{sm}.

\begin{figure*}
	\includegraphics[width=\linewidth]{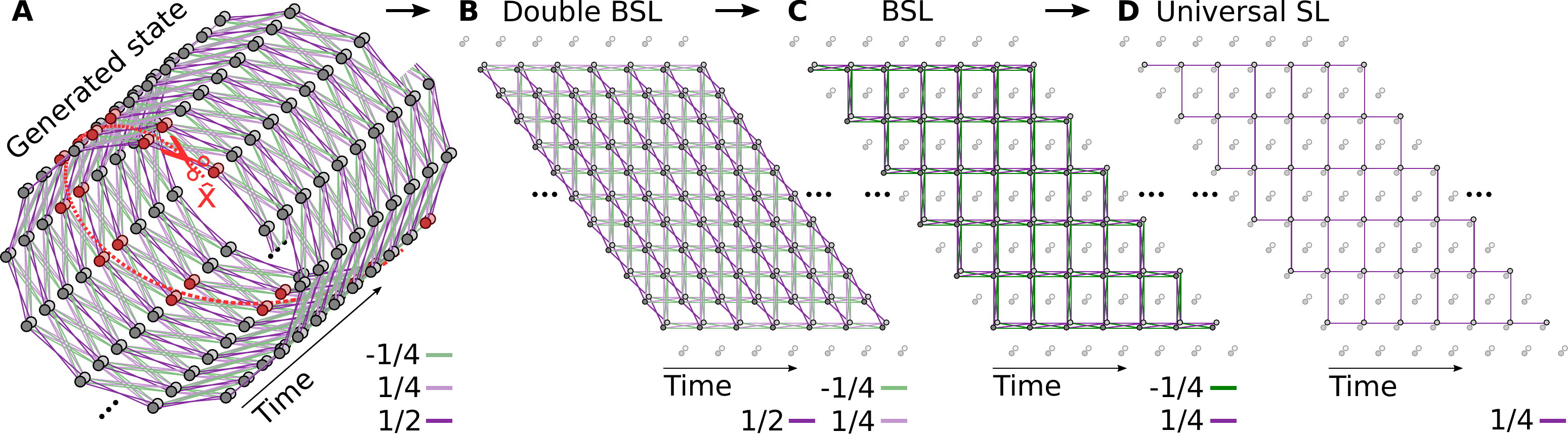}
	\caption{Universality of generated 2D cluster state. (A) Graph of the generated 2D cluster state. Measuring the nodes marked by red in the position basis removes all edges connected to the measured nodes, and the cylindrical graph unfolds to a plane. (B) Resulting plane 2D cluster state after the projective measurements in (A), consisting of two  bilayer square lattices (double BSL) connected by edges of weight 1/2. (C) Single BSL after projective measurement of half the modes in (B) in the position basis. (D) Square lattice (SL) after projective position measurements of all modes in spatial mode $B$ (dark nodes), and applying the Fourier gate ($\pi/2$ phase delay) on half the modes in spatial mode $A$ (bright nodes). This SL is a traditional universal resource state for MBQC.}
\end{figure*}

The produced cylindrical 2D cluster state can be shown to be a universal resource for quantum computing: In Fig. 2, the generated cylindrical cluster state is unfolded and projected into a square lattice by projective measurements in the position basis and $\pi/2$ phase-space rotations of different modes. Such a square lattice is a well-known universal resource for quantum computing \cite{gu09}, and thus the initial cylindrical cluster state is itself universal. For computation it is not necessary to project the generated cluster state into a square lattice---rather, one would in general optimize the detector settings required for the gate to be implemented. For instance, with proper settings the cluster state can be projected into 1D dual-rail wires along the cylinder, an efficient resource for one-mode computation \cite{yokoyama13,alexander18} and with possible two-mode interactions between them---for details see Supplementary Text section 1.4 \cite{sm}. Doing so requires fast control of the measurement bases in between temporal modes, while in this work the cluster state is measured in fixed bases for state verification.

Multi-partite cluster state inseparability can be witnessed through the measurement of the uncertainties of the state nullifiers---linear combinations of position and momentum operators for which the cluster states are eigenstates with eigenvalue 0. E.g. for the ideal two-mode EPR state, the well-known nullifiers are $\hat{n}^x_\text{EPR}=\hat{x}_A-\hat{x}_B$ and $\hat{n}^p_\text{EPR}=\hat{p}_A+\hat{p}_B$ since $\hat{n}^x_\text{EPR}\ket{\rm{EPR}}=0$ and $\hat{n}^p_\text{EPR}\ket{\rm{EPR}}=0$. For our 2D cluster state, $|\text{2D}\rangle$, the nullifiers consist of 8 modes and are given by 
\begin{equation}\begin{split}\label{eq:nullifierX}
	\hat{n}_k^x = \;&\hat{x}_{k}^A + \hat{x}_{k}^B - \hat{x}_{k+1}^A - \hat{x}_{k+1}^B\\
	& \quad- \hat{x}_{k+N}^A + \hat{x}_{k+N}^B - \hat{x}_{k+N+1}^A + \hat{x}_{k+N+1}^B\;,
\end{split}\end{equation}
\begin{equation}\begin{split}\label{eq:nullifierP}
	\hat{n}_k^p = \;&\hat{p}_{k}^A + \hat{p}_{k}^B + \hat{p}_{k+1}^A + \hat{p}_{k+1}^B\\
	&\quad - \hat{p}_{k+N}^A + \hat{p}_{k+N}^B + \hat{p}_{k+N+1}^A - \hat{p}_{k+N+1}^B\;,
\end{split}\end{equation}
as $\hat{n}_k^x|\text{2D}\rangle=0$ and $\hat{n}_k^p|\text{2D}\rangle=0$ (derived in the Supplementary Text section 1.3 \cite{sm}), where the subscript indicates the temporal mode index with $N$ being the number of temporal modes in the cluster state circumference.

\begin{figure}
	\includegraphics[width=\linewidth]{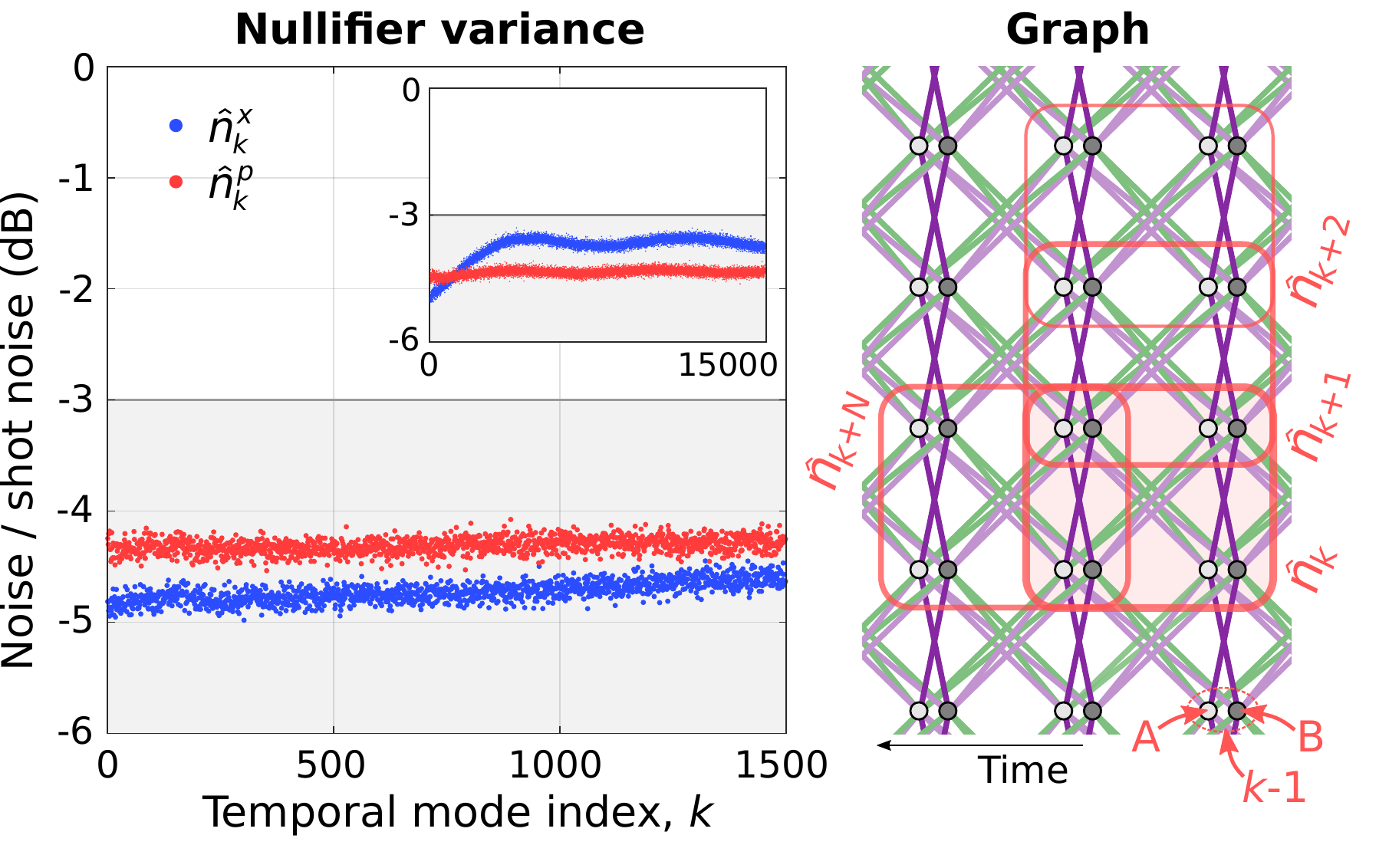}
	\caption{Experimental result. On the right graph, the nullifiers in eq.~(\ref{eq:nullifierX}) and (\ref{eq:nullifierP}) are shown on the 2D cluster state lattice with the measured variance of $\SI{1500}{}$ consecutive nullifiers shown in the left plot. Here, the variance is calculated from $\SI{10000}{}$ measurements of each nullifier.  All nullifier variances are seen to be well below the $\SI{-3}{dB}$ inseparability bound derived in the Supplementary Text section 2 \cite{sm}, and thus the generated cluster state is completely inseparable. In the insert, the nullifier variance of a larger data set with $2\times\SI{15000}{}=\SI{30000}{}$ modes are shown. Again, with all modes below the $\SI{-3}{dB}$ inseparability bound, we conclude the successful generation of a $\SI{30000}{}$ mode 2D cluster state. The rapid increase of the variance in $\hat{n}_k^x$ and its periodic variation is caused by phase fluctuation of the squeezing sources as described in the Supplementary Text section 4 \cite{sm}. }
\end{figure}

The practically realizable cluster state is never an exact eigenstate of the nullifiers since such a state is unphysical. The measurement outcomes of the nullifiers are therefore not exactly zero in every measurement but possess some uncertainties around zero. A condition for complete inseparability of the 2D cluster state (derived in the Supplementary Text section 2 \cite{sm}) leads to a bound on the variances of all nullifiers of $\SI{3}{dB}$ squeezing below the shot noise level. Therefore, to witness full inseparability, we must observe more than $\SI{3}{dB}$ squeezing for all nullifiers. In Fig.~3, the measured nullifier variances are shown for a dataset of $\SI{1500}{}$ nullifiers and they are all observed to be well below the $\SI{-3}{dB}$ bound; we measure an averaged variance of $\SI{-4.7}{dB}$ and $\SI{-4.3}{dB}$ for $\hat{n}_k^x$ and $\hat{n}_k^p$, respectively. In the inset of Fig. 3, we present the measurement of a longer cluster state of $\SI{15000}{}$ temporal modes corresponding to a measurement time of $\SI{4}{ms}$. Although phase instabilities are clearly seen to affect the performance in terms of variations of the nullifier variances, all variances stay below the $\SI{-3}{dB}$ bound. The 2D cluster of $2\times \SI{15000}{} = \SI{30000}{}$ modes is thus fully inseparable. Note that not all $\SI{30000}{}$ modes of the cluster state need to exist simultaneously when performing projective measurements for computation. In fact, only a single temporal mode of the cluster state needs to exist while the remaining modes of the state are under construction. Hence, the cluster state can be immediately consumed for computation while being generated, with no additional state storage necessary---see Supplementary Text section 1.4 for a possible measurement scheme for computation on the cluster state.

With the deterministic generation of a universal 2D cluster state, we have for the first time (in parallel with Asavanant et al. \cite{asavanant19}) in any system constructed a platform for universal MBQC. Its scalability was demonstrated by entangling $\SI{30000}{}$ optical modes in a 2D lattice that includes 24 input modes and allows for a computation depth of $\SI{1250}{}$ modes. Since only a few modes exist simultaneously, we are not limited by the coherence time of the light source, and thus the number of operations depends only on the phase stability of the system. The computational depth can therefore be unlimited by implementing continuous feedback control of the system for phase stabilization as demonstrated for the 1D photonic cluster state in \cite{yoshikawa16}. The results presented here and in \cite{asavanant19} are similar: Both 2D cluster states are generated deterministically in the CV regime with comparable size and amount of squeezing in the nullifier variance. However, with only two squeezing sources, three interference points, and operation in fiber, the experimental setup demonstrated here is simpler, while in \cite{asavanant19} larger bandwidth OPOs are demonstrated resulting in shorter delay lines. In both systems, the number of input modes can be readily increased by using OPOs with larger bandwidths, possibly combined with a longer time delay of the second interferometer. E.g. using OPOs with a $\SI{1}{GHz}$ bandwidth (65 times wider) and a twice as long interferometer delay, a state with $\sim\SI{1500}{}$ input modes can be generated. Large bandwidth OPOs have been demonstrated, but phase stability and losses in the delay lines are more challenging. While phase fluctuation is only a matter of experimental control on which we expect to improve with continuous phase stabilization, delay losses are unavoidable and increasing the OPO bandwidth may be a better solution than increasing the delay lengths.
 
CV cluster states are described by Gaussian statistics, but it is known that an element (state, operation, or measurement) of non-Gaussian quadrature statistics is required for universal quantum computing \cite{bartlett02}. Such an element could be a photon number resolving detector (PNRD) or an ancillary cubic-phase state \cite{alexander18,gottesman01}. Despite recent experimental efforts in developing high-efficiency PNRD \cite{thekkadath19} and deterministically generating optical states with non-Gaussian statistic \cite{hacker19}, the formation of the required non-Gaussianity of the cluster state still constitutes an important challenge to be tackled in the future. Another currently limiting factor towards quantum computation is the existence of finite squeezing in the cluster leading to excess quantum noise and thus computational errors. However, these errors can be circumvented using Gottesman-Kitaev-Preskill (GKP) state encoding \cite{gottesman01} concatenated with traditional qubit error correction schemes leading to fault-tolerant computation with a 15--$\SI{17}{dB}$ squeezing threshold \cite{walshe19}. Another recently discovered advantage of the GKP encoding is that in addition to fault-tolerance, it also allows for universality without adding extra non-Gaussian states or operations \cite{baragiola19}. While GKP states have recently been produced in the microwave regime \cite{campagne19} and in trapped-ion mechanical oscillators \cite{fluhmann19}, their production in the optical regime remains a task for future work.  For further discussion on quantum computation using the generated cluster state, see Supplementary Text section 1.4 \cite{sm}. Although a path towards fault-tolerant universal quantum computing using CV cluster states has been established, it is highly likely that the first demonstrations of CV quantum computation will be non-universal algorithmic sub-routines such as boson sampling and instantaneous quantum computing \cite{su18}. With the large, but noisy cluster state demonstrated here, interesting future work will be to implement basic Gaussian circuits and investigate e.g. the attainable circuit depth. Furthermore, the technique of folding a 1D cluster state into a 2D structure could be extended, using an additional interferometer, to form 3D cluster states which might be suitable for topologically protected MBQC.

\section*{Acknowledgements}
We thank R. N. Alexander for useful discussion on the final manuscript and J. B. Brask for proofreading. The work was supported by the Danish National Research Foundation through the Center for Macroscopic Quantum States (bigQ, DNRF142), and the VILLUM FOUNDATION Young Investigator Programme (grant no. 10119).

{\bf Author contributions:} M.V.L. and U.L.A. conceived the project. J.S.N., X.G. and C.R.B. designed and built the squeezing sources, while M.V.L. developed the theoretical background, designed the experiment and built the setup. M.V.L. performed the experiments and data analysis. The project was supervised by U.L.A. and J.S.N. The manuscript was written by U.L.A., M.V.L. and J.S.N. with feedback from all authors.

{\bf Data availability:} Experimental data and analysis code is available at figshare \cite{data}.

%%% References %%%

\end{document}

\noindent{\LARGE\textbf{Material and Methods}}
\section*{Experimental design}
\begin{figure}[b!]
	\centering
	\includegraphics[width=0.9\textwidth]{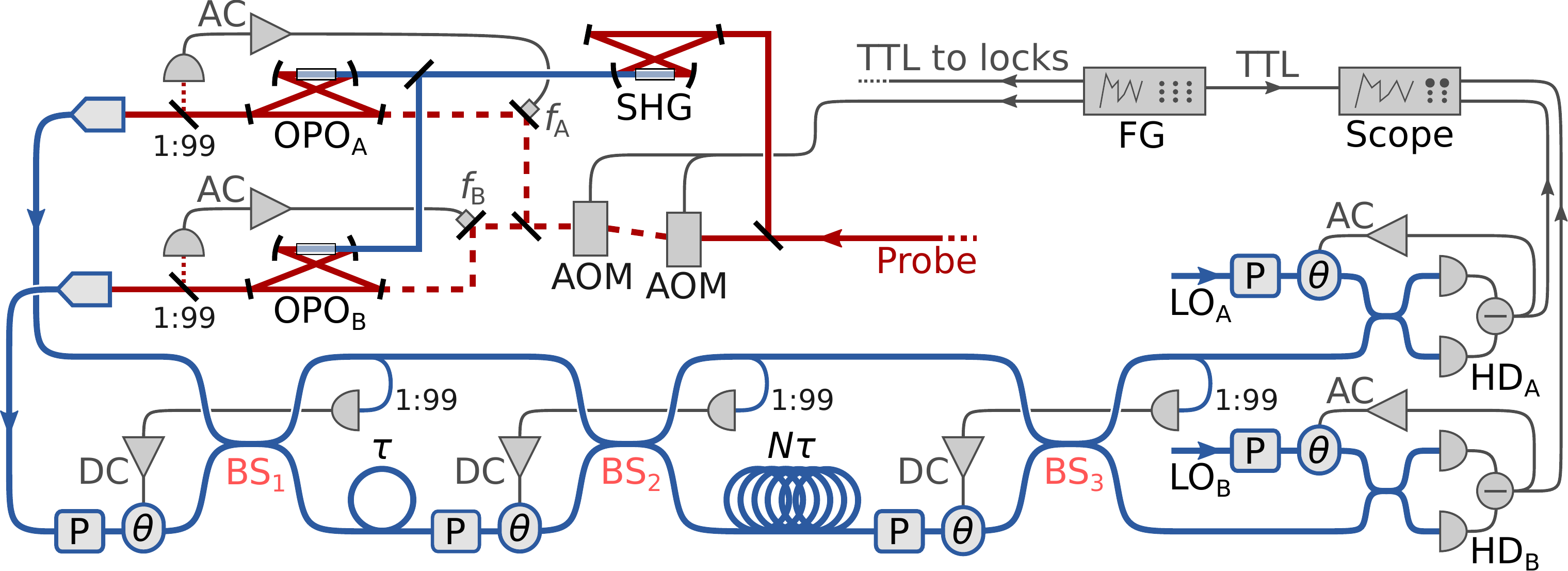}
	\caption{Detailed schematic of the experimental setup for 2D cluster state generation. Here the free space squeezing sources are marked by red (besides second harmonic generated light at $\SI{775}{nm}$ wavelength which is marked by blue), while optical fibers in which the cluster state is generated are marked by blue. Electronics for experimental control are marked by black. A function generator (FG) generates a logic signal (TTL) for switching on and off the probe and activating/deactivating feedback for cavity and phase locks. Data is acquired on an oscilloscope (Scope) when the probe is turned off and feedback is kept constant. The fiber components marked by P and $\theta$ represents manual polarization controllers and phase control by fiber-stretchers respectively.}
	\label{fig:detailed_setup}
\end{figure}

The experimental setup is shown in detail in Fig.~\ref{fig:detailed_setup}. Amplitude squeezed light at $\SI{1550}{nm}$ wavelength is generated by type-0 parametric down conversion in two bow-tie shaped optical parametric oscillators ($\text{OPO}_\text{A}$ and $\text{OPO}_\text{B}$) with periodically poled potassium titanyl phosphate (PPKTP) crystals, pumped by light at $\SI{775}{nm}$ wavelength generated from a second harmonic generator (SHG). For cavity and phase locking throughout the setup, we use a sample-hold locking scheme where the two OPOs are periodically seeded with a coherent probe chopped by two acousto-optic modulators (AOM): During the sample-time the probe is left on and active feedback is used for cavities and phase locks. After $\SI{10}{ms}$ of sample-time with active feedback, the probe is turned off for $\SI{5}{ms}$ (denoted hold-time) where all feedback loops are kept constant and quadrature data of the generated 2D cluster state is acquired from the two homodyne detectors (HD$_\text{A}$ and HD$_\text{B}$). The cavities are locked by the Pound-Drever-Hall locking technique using a counter propagating lock beam with $\SI{28}{MHz}$ phase modulation by an electro-optic modulator (not shown in Fig.~\ref{fig:detailed_setup}). For the generation of amplitude squeezing, the classical parametric gains in $\text{OPO}_\text{A}$ and $\text{OPO}_\text{B}$ are locked to de-amplification using an AC-locking scheme: Phase modulated probe beams (at frequencies $f_A=\SI{90}{kHz}$ and $f_B=\SI{55}{kHz}$) are injected into the OPOs, a fraction (1\%) is measured and subsequently fed back to piezoelectric mounted mirrors.

The beams of squeezed light are coupled into single mode fibers (SMF) using gradient-index (GRIN) lenses with 97\% coupling efficiency. Here, the two beams of squeezed light are interfered in a 50:50 fiber coupler ($\text{BS}_1$), where 1\% of one output arm is tapped, detected, and fed back to a phase controlling fiber-stretcher for locking the relative phase between the two input beams. For more information on this fiber-stretcher, see previous experimental work in {\citeLarsen}. Using a manual polarization controller, the visibility is optimized to near unity. By locking the relative phase difference to $\pi/2$ using a DC-locking scheme, EPR-states are generated.

Using a short delay line consisting of $\SI{50.5}{m}$ SMF-28e+ fiber, one spatial mode is delayed by $\tau=\SI{247}{ns}$. This delay defines the temporal mode width. Again, the two spatial modes are interfered on a 50:50 fiber coupler ($\text{BS}_2$) with phase control by tapping and detecting 1\% of the output and feeding back to a fiber-stretcher, while visibility is optimized with a manual polarization controller. Locking the phase with a DC-locking scheme leads to a 1D cluster state with temporal modes defined by the short $\tau$-delay.

Finally, using a long delay of $\SI{606}{m}$, one spatial mode is delayed by $N=12$ temporal modes. Interfering the two spatial mode in the 50:50 fiber coupler ($\text{BS}_3$) corresponds to "coiling up" the 1D cluster state generated in $\text{BS}_2$, leading to a 2D cluster state as illustrated in the main text Fig.~\setup \ and described in the Supplementary Text section \ref{sec:2Dcluster}. Here, too, the relative phase is locked by tapping and detecting 1\% of the output and feeding back to a fiber-stretcher, while polarization is controlled with a manual polarization controller.

For characterizing the generated 2D cluster state, amplitude ($\hat{x}$) and phase ($\hat{p}$) quadratures of the two spatial modes are continuously measured by two fiber-based homodyne detectors (HD). For more information on these fiber-based HDs, see previous experimental work in {\citeLarsen}. The local oscillator phases for the two HDs are locked using an AC-locking scheme, where for measuring in the $\hat{x}$- and $\hat{p}$-basis, demodulation by $f_A$ and $f_B$ are used, respectively.

For more details and characterization of the experimental implementation given here, see Supplementary Text section \ref{sec:expMethods}.

\newpage
\noindent{\LARGE\textbf{Supplementary Text}}
	
\section{Theory on cluster state}
In this section, cluster states are first introduced in section \ref{sec:clusterIntro} before the generated 2D cluster state is derived in section \ref{sec:cluster_state_generation} and its nullifiers in section \ref{sec:nullifiers}. We use the convention of $\hbar=1$.

\subsection{Introduction}\label{sec:clusterIntro}
Cluster states are a resource for measurement based quantum computation (MBQC) and are well described in {\citeGu} for the case of continuous variables (CV). For CV a cluster state is a set of modes, all initially in the momentum eigenstate $\ket{0}_p$, entangled by a number of controlled-Z operations of weight $g$, $\hat{C}_Z=\exp\left[ig\hat{x}\otimes \hat{x}\right]$ where $\hat{x}$ is the position quadrature. In the following we follow the conventions of graphical calculus for Gaussian pure states outlined in {\citeMenicucci}, and more details on the theory summarized here can be found in {\citeMenicucciGu}.

\begin{figure}[b]
	\centering
	\includegraphics[width=0.7\textwidth]{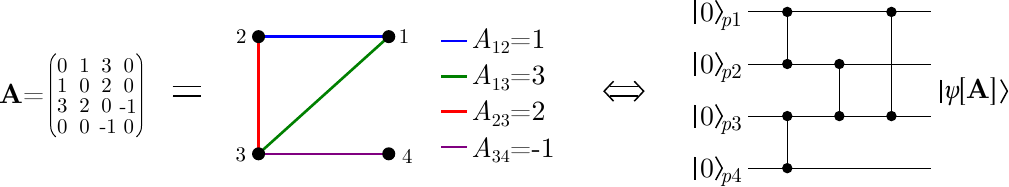}
	\caption{Adjacency matrix with its corresponding graph and equivalent circuit model.}
	\label{fig:A}
\end{figure}

A cluster state $\ket{\psi_\textbf{A}}$ of $m$ modes can be defined by a symmetric real valued $m\times m$ adjacency matrix $\textbf{A}$ as
\begin{equation}\label{eq:A}
	\ket{\psi_\textbf{A}}=\hat{C}_Z[\textbf{A}]\ket{0}_p^{\otimes m}=\prod_{j=1}^{m}\prod_{k=j}^{m}e^{iA_{jk}\hat{x}_j \hat{x}_k}\ket{0}_p^{\otimes m}=\exp\left[\frac{i}{2}\hat{\textbf{x}}^\text{T}\textbf{A}\hat{\textbf{x}}\right]\ket{0}_p^{\otimes m}\;,
\end{equation}
where $\hat{\textbf{x}}=(\hat{x}_1,\hat{x}_2,\cdots,\hat{x}_m)^\text{T}$ is vector of position operators. For ideal cluster states, \textbf{A} is zero in the diagonal, while the off-diagonal term $A_{jk}$ describes a link (an edge) between mode $j$ and $k$ by the $\hat{C}_Z$-operator of weight $A_{jk}$. We can picture a cluster state as a graph from its adjacency matrix as in Fig.~\ref{fig:A}.

The cluster state in eq.~(\ref{eq:A}) is most easily described in the stabilizer formalism in which $\hat{p}_j-\sum_kA_{jk}\hat{x}_k$ is a nullifier:
\begin{equation}\label{eq:null}
	\left(\hat{\textbf{p}}-\textbf{A}\hat{\textbf{x}}\right)\ket{\psi_\textbf{A}}=\textbf{0}\;,
\end{equation}
where $\hat{\textbf{p}}=(\hat{p}_1,\hat{p}_2,\cdots,\hat{p}_m)^\text{T}$ is a vector of momentum operators. In conclusion, when measuring the nullifier $\hat{p}_j-\sum_kA_{jk}\hat{x}_k$ we expect vanishing variance. $\textbf{A}$ gives a complete description of the state $\ket{\psi_\textbf{A}}$.

\subsubsection{Approximate cluster states}
Eq.~(\ref{eq:null}) is only valid for true momentum eigenstates as in eq.~(\ref{eq:A}), which require infinite squeezing and are not physical. Finite squeezing leads to non-zero variance when measuring the nullifier, and the variance increases with decreasing squeezing. Finite squeezing can be accounted for in the adjacency matrix by allowing it to be complex. We denote this complex adjacency matrix
\begin{equation*}
	\textbf{Z}=\textbf{V}+i\textbf{U}\;,
\end{equation*}
where \textbf{V} and \textbf{U} are real valued and symmetric. Again, \textbf{V} is zero in its diagonal and corresponds to \textbf{A} in the ideal case, while most often \textbf{U} is non-zero in the diagonal and corresponds to the deviation from the ideal case. We can still illustrate the corresponding graph state as in Fig.~\ref{fig:A}, but with complex weight and with self-loops on each node corresponding to the imaginary non-zero diagonal terms of \textbf{Z}.

The physical graph state described by \textbf{Z} is said to be an approximate cluster state with adjacency matrix \textbf{A} if
\begin{equation*}
	\lim_{r\rightarrow\infty} \textbf{Z}(r) = \textbf{A}\;,
\end{equation*}
where $r$ is the squeezing parameter of the initial states. As an example, applying $\hat{C}_Z[\textbf{A}]$ to a number of finitely squeezed momentum states leads to
\begin{equation*}
	\textbf{Z}=\textbf{A}+ie^{-2r}\textbf{I} \rightarrow \textbf{A} \; \text{for}\; r\rightarrow\infty\;.
\end{equation*}
Here $\textbf{V}=\textbf{A}$ and $\textbf{U}=e^{-2r}\textbf{I}$.

\subsubsection{$\mathcal{H}$-graph states}
The controlled-Z operation, $\hat{C}_Z$, for entanglement generation is not easily implemented experimentally. Instead, quadrature entanglement (two-mode squeezing) is generated directly by non-degenerate down conversion or by interference of squeezed states, and the resulting graph state can be expressed by the adjacency matrix
\begin{equation}\label{eq:Hgrapoh}
	\textbf{Z}=ie^{-2r\textbf{G}}\;
\end{equation}
where \textbf{G} is a real symmetric matrix. The state is called an $\mathcal{H}$-graph state, since it can be generated by the Hamiltonian
\begin{equation}\label{eq:Ham}
	\hat{\mathcal{H}}(\textbf{G})=\hbar\kappa\left(\hat{\textbf{x}}^\text{T}\textbf{G}\hat{\textbf{p}}+\hat{\textbf{p}}^\text{T}\textbf{G}\hat{\textbf{x}}\right)\;,
\end{equation}
with $\kappa$ being the squeezing parameter per unit time, $r=2\kappa t$. It is not easy to illustrate this graph state with its exponential map, but in the case of \textbf{G} being self-inverse ($\textbf{G}^2=\textbf{I}$), eq.~(\ref{eq:Hgrapoh}) simplifies to
\begin{equation}\label{eq:Zsi}
	\textbf{Z}=i\cosh(2r)\textbf{I}-i\sinh(2r)\textbf{G}\;,
\end{equation}
and it can be pictured as in Fig.~\ref{fig:A} with complex weights. However, it is not an approximate cluster state as \textbf{Z} does not go to some real valued matrix with zero in the diagonal for $r\rightarrow\infty$. But in the case of \textbf{G} also being bipartite (meaning the nodes can be separated into two sets with no connecting edges in between modes of the same set), it can be transformed into an approximate cluster state by applying the Fourier gate ($\pi/2$ rotation in phase-space) on some of its modes. Finally, since this Fourier gate can be absorbed into the measurement basis when measuring each mode of the graph state, we consider generation of a self-inverse bipartite $\mathcal{H}$-graph state as cluster state generation.

\subsection{Cluster state generation}\label{sec:cluster_state_generation}
In the approach to cluster state generation, we start with modes of quadrature squeezed light to which we apply beam-splitters and Fourier gates. Traditionally, the starting point is the complex adjacency matrix for $m$ modes squeezed in the phase (or momentum) quadrature,
\begin{equation}\label{eq:Zsq}
	\textbf{Z}=ie^{-2r}\textbf{I}\;,
\end{equation}
with $r$ being the squeezing parameter. In the experimental implementation we start with states squeezed in the amplitude (or position) quadrature, but this makes no difference to the theoretical derivation of the cluster state, and is merely a question on quadrature definition or $\pi/2$ phase-space rotation. The quadrature transformation under beam-splitter transformations and/or phase-space rotations in the Heisenberg picture can be expressed by a $2m\times2m$ symplectic matrix \textbf{S} as
\begin{equation*}
	\begin{pmatrix}
		\hat{\textbf{x}}'\\\hat{\textbf{p}}'
	\end{pmatrix}
	=\textbf{S}
	\begin{pmatrix}
		\hat{\textbf{x}}\\\hat{\textbf{p}}
	\end{pmatrix}\quad,\quad \textbf{S}=
		\begin{pmatrix}
		\textbf{A} & \textbf{B}\\\textbf{C} & \textbf{D}
	\end{pmatrix}\;,
\end{equation*}
where \textbf{A}, \textbf{B}, \textbf{C} and \textbf{D} are real $m\times m$ matrices. The corresponding transformation of the adjacency matrix \textbf{Z} is shown in {\citeMenicucci} to be
\begin{equation}\label{eq:newZ}
	\textbf{Z}'=\left(\textbf{C}+\textbf{D}\textbf{Z}\right)\left(\textbf{A}+\textbf{B}\textbf{Z}\right)^{-1}\;,
\end{equation}
with the resulting graph described by $\textbf{Z}'$.

\begin{figure}
	\centering
	\includegraphics[width=0.9\textwidth]{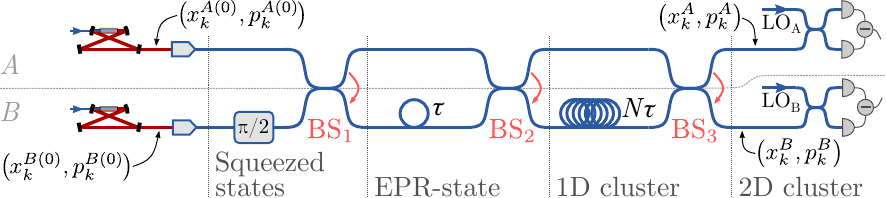}
	\caption{Sketch of setup for 2D cluster state generation. Following {\citeYokoyama}, first a 1D cluster state ($\mathcal{H}$-graph state) is generated with temporal modes separated by the time $\tau$ using beam-splitters $\text{BS}_1$ and $\text{BS}_2$ together with the optical delay $\tau$. This 1D cluster state is then coiled up in a cylinder with the $N\tau$ delay, such that temporal modes at times $k\tau$ in the spatial mode $A$ overlap in time with the temporal modes of initial times $(k-N)\tau$ in the spatial $B$, where $k$ is an integer. From the side of the cylinder, we can see it as parallel 1D cluster states, which are then connected by the last beam-splitter $\text{BS}_3$ to form a 2D cylindrical cluster state. The arrows on the beam-splitters points from the first to the second mode of the beam-splitter transformation $\textbf{S}_{\text{BS}}^{AB}$ in eq.~(\ref{eq:Soperations}).}
	\label{fig:setup}
\end{figure}

The scheme of 2D cluster state generation in the main text Fig.~\setup {} is summarized in Fig.~\ref{fig:setup}. First a 1D $\mathcal{H}$-graph state is generated as in {\citeYokoyama}, by applying a $\pi/2$ phase-space rotation in the spatial mode $B$, beam-splitter transformation, delay of one spatial mode and another beam-splitter transformation. The phase-space rotation and beam-splitter is described by symplectic operations, while the delay is included by keeping track of the temporal mode index of simultaneously existing temporal modes in the two spatial modes $A$ and $B$. In the following sections, each step is described in detail.

\subsubsection{EPR-state generation}\label{sec:EPR}
As the first step in Fig.~\ref{fig:setup}, consider two modes $A$ and $B$ squeezed in the phase quadratures. To generate an EPR-state, mode $B$ is rotated by $\pi/2$ in phase-space, and we apply the beam-splitter transformation $\text{BS}_1$ between $A$ and $B$. The symplectic matrix is
\begin{equation}\label{eq:Soperations}
	\textbf{S}=\textbf{S}^{AB}_{\text{BS}}\textbf{S}^B_{\pi/2}\quad,\quad
	\textbf{S}^B_{\pi/2}=
	\begin{pmatrix}
		1 & 0 & 0 & 0\\
		0 & 0 & 0 & -1\\
		0 & 0 & 1 & 0\\
		0 & 1 & 0 & 0
	\end{pmatrix}\quad,\quad
	\textbf{S}_{\text{BS}}^{AB}=\frac{1}{\sqrt{2}}
	\begin{pmatrix}
		1 & -1 & 0 & 0\\
		1 & 1 & 0 & 0\\
		0 & 0 & 1 & -1\\
		0 & 0 & 1 & 1
	\end{pmatrix}\;.
\end{equation}
Identifying \textbf{A}, \textbf{B}, \textbf{C} and \textbf{D} in (\ref{eq:newZ}) from (\ref{eq:Soperations}) and inserting (\ref{eq:Zsq}) we get
\begin{equation}\label{eq:Zepr}
	\textbf{Z}_\text{EPR}=\begin{pmatrix}
		i\cosh(2r)&-i\sinh(2r)\\-i\sinh(2r)&i\cosh(2r)
	\end{pmatrix}\;,
\end{equation}
which is an $\mathcal{H}$-graph with the exact form of (\ref{eq:Zsi}) where
\begin{equation*}
	\textbf{G}=\begin{pmatrix}
	0&1\\1&0\
	\end{pmatrix}\;.
\end{equation*}
Note that the same EPR-state can then be generated by the Hamiltonian in (\ref{eq:Ham}), corresponding to non-degenerate parametric down conversion as expected. \textbf{G} is self-inverse and bipartite, and if we were to rotate mode $B$ (applying $\textbf{S}^B_{\pi/2}$) we would get
\begin{equation*}
	\textbf{Z}_\text{EPR}'=\begin{pmatrix}
		i\,\text{sech}(2r) & \tanh(2r)\\ \tanh(2r) &i\,\text{sech}(2r)
	\end{pmatrix}
	\rightarrow\begin{pmatrix}
	0 & 1\\ 1& 0
	\end{pmatrix}\equiv\textbf{A}
	\; \text{for}\;r\rightarrow\infty\;,
\end{equation*}
and so the $\mathcal{H}$-graph for the EPR-state has a corresponding approximate cluster state. From eq.~(\ref{eq:null}), the nullifiers of this cluster state are $\hat{p}_A-\hat{x}_B$ and $\hat{p}_B-\hat{x}_A$, which transform into $\hat{p}_A+\hat{p}_B$ and $\hat{x}_B-\hat{x}_A$ after rotating mode $B$ by $\pi/2$. These relations are expected for an EPR-state.

\subsubsection{1D cluster states}\label{sec:1Dcluster}
To generate 1D cluster states as in {\citeYokoyama}, we continue with pairs of EPR-states as described by the adjacency matrix in eq.~(\ref{eq:Zepr}). Instead of the matrix notation, we will use the more convenient graph notation:
\begin{center}
	\includegraphics[width=0.6\textwidth]{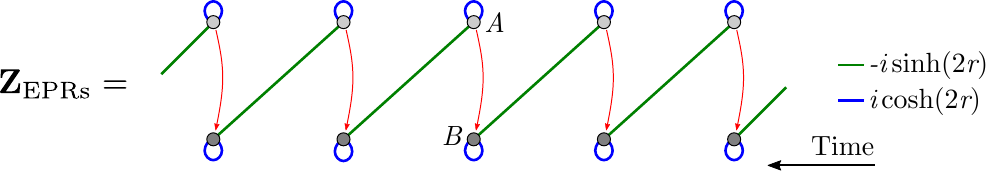},
\end{center}
with the beam-splitter transformation marked by red arrows corresponding to $\text{BS}_2$ in Fig.~\ref{fig:setup}. Here, the bright and dark grey nodes symbolize temporal modes of the two different spatial modes of $A$ and $B$ respectively, and has no other meaning than distinguishing spatial modes. Note also that $\textbf{Z}_\text{EPRs}$ is the graph just after the delay, $\tau$, in Fig.~\ref{fig:setup}. After the beam-splitter transformation connecting the pairs of EPR-states, we attain the 1D $\mathcal{H}$-graph state
\begin{center}
	\includegraphics[width=0.6\textwidth]{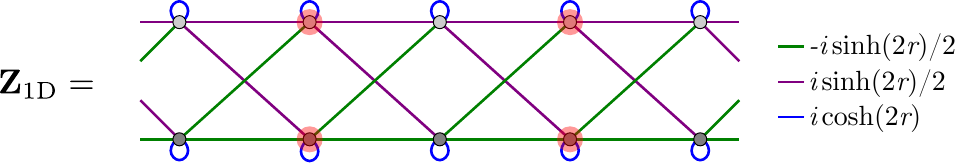},
\end{center}
which is self-inverse and bipartite, and so it can be transformed into an approximative cluster state by applying the Fourier gate on all modes in one of the bipartitions: Rotating every second pairs of spatial modes marked with red in $\textbf{Z}_\text{1D}$ leads to 
\begin{center}
	\includegraphics[width=0.6\textwidth]{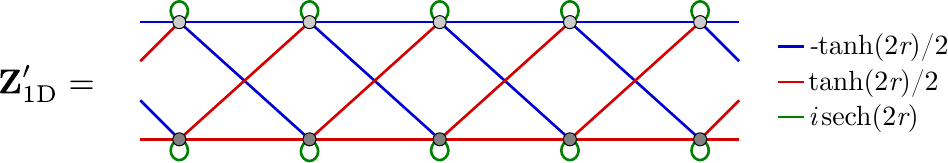},
\end{center}
with only real edges and vanishing self-loops when $r\rightarrow\infty$ as $\tanh(2r)/2\rightarrow 1/2$ and $i\,\text{sech}(2r)\rightarrow0$.
By determining the nullifiers in the limit $r\rightarrow\infty$, and rotating every second pair of modes back again (as for the EPR-state in section \ref{sec:EPR}) we can determine the nullifiers of $\textbf{Z}_\text{1D}$, which each will include 5 modes according to eq.~(\ref{eq:null}) (all modes connected to a single mode). These nullifiers can be simplified, as all linear combinations of nullifiers are also nullifiers, and the nullifiers including the least modes are
\begin{equation*}
	\hat{x}_{Ak}+\hat{x}_{Bk}-\hat{x}_{Ak+1}+\hat{x}_{Bk+1}\quad,\quad \hat{p}_{Ak}+\hat{p}_{Bk}+\hat{p}_{Ak+1}-\hat{p}_{Bk+1}\;,
\end{equation*}
where the index $k$ and $k+1$ denote different temporal mode numbers. Since the nullifiers are linear combination of $\hat{x}$ \textit{or} $\hat{p}$, they are easily measured in order to verify the entanglement of the cluster state.

\subsubsection{2D cluster states}\label{sec:2Dcluster}
\begin{figure}
	\centering
	\includegraphics[width=0.5\textwidth]{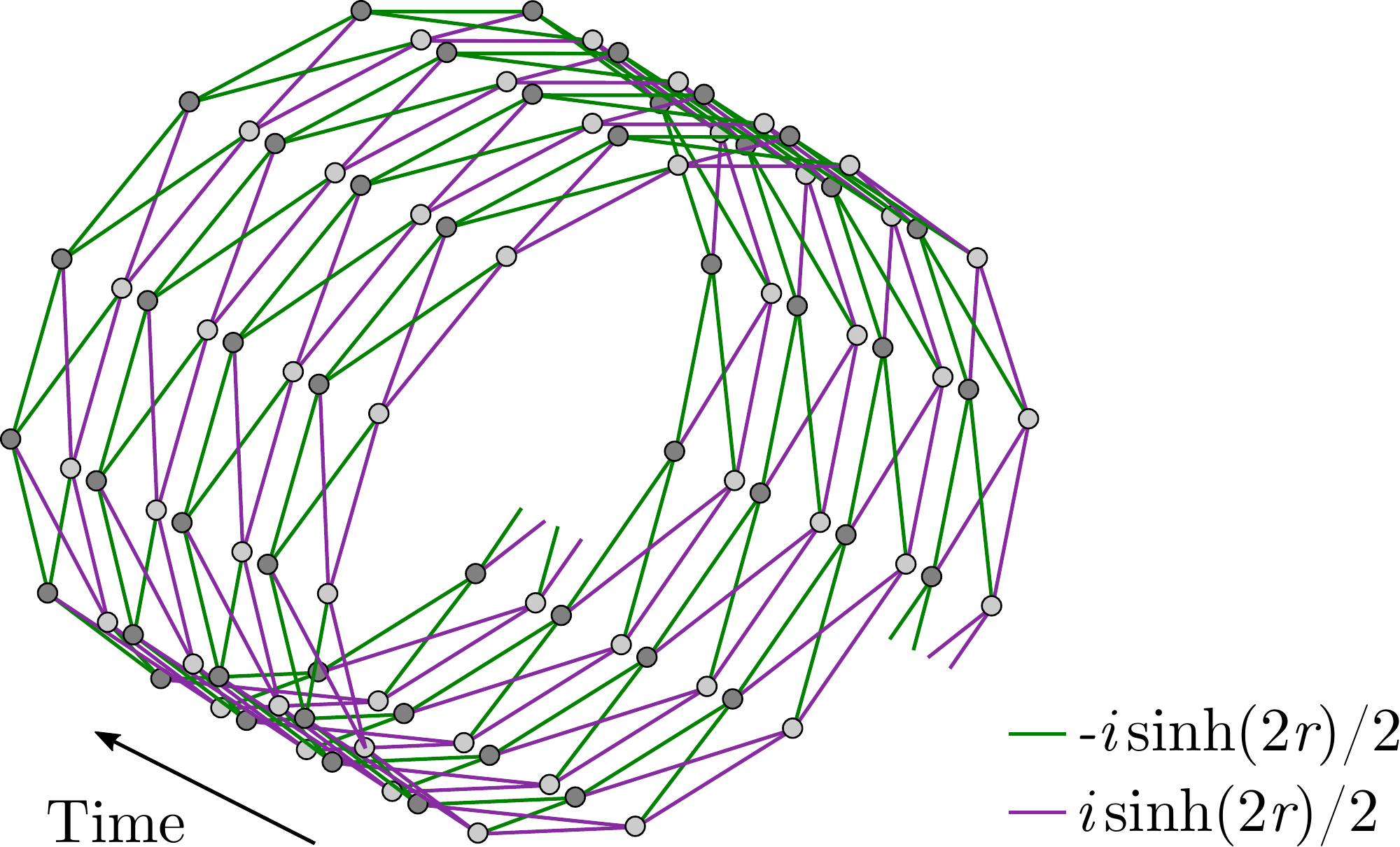}
	\caption{Complex adjacency matrix, $\textbf{Z}_\text{1Ds}$, of the coiled up 1D $\mathcal{H}$-graph state just after the $N\tau$ delay in Fig.~\ref{fig:setup} with $N=12$ as in the experimental implementation. For simplicity, self-loops of $i\cosh(2r)$ are omitted, but they are still present in the diagonal of $\textbf{Z}_\text{1Ds}$.}
	\label{fig:cylinder}
\end{figure}

After the $N\tau$ delay in Fig.~\ref{fig:setup}, the 1D cluster, $\textbf{Z}_\text{1D}$, is coiled up into a cylinder as illustrated in Fig.~\ref{fig:cylinder}. To begin with, we consider only a section of the cylinder:
\begin{center}
	\includegraphics[width=0.6\textwidth]{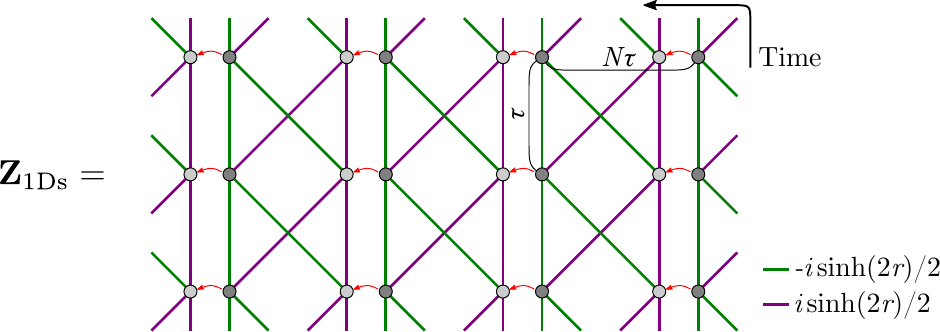},
\end{center}
where each parallel 1D cluster state is separated by $N\tau$ in time corresponding to one circumference of the cylinder. Note that the self-loops of $i\cosh(2r)$ have been omitted, and will be omitted in the following, but they are still present in the diagonal of $\textbf{Z}_\text{1Ds}$. Here, two closer spaced spatial modes $A$ and $B$ overlap in time, and the red arrows represent the last beam-splitter transformation $\text{BS}_3$ in Fig.~\ref{fig:setup}, leading to the 2D $\mathcal{H}$-graph state
\begin{center}
	\includegraphics[width=0.6\textwidth]{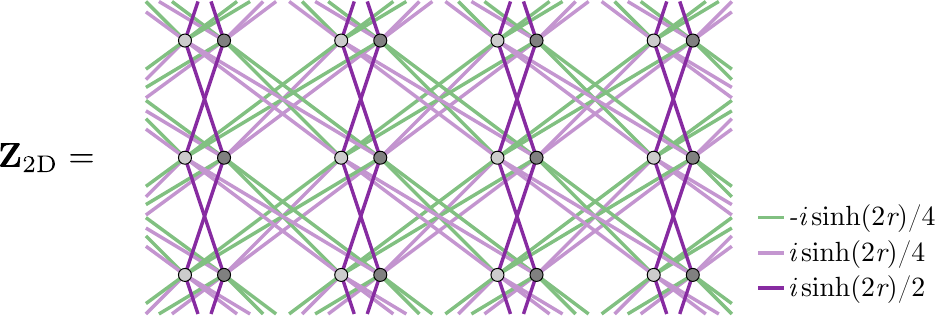}.
\end{center}
$\textbf{Z}_\text{2D}$ is self-inverse, and if we consider $\textbf{Z}_\text{2D}$ as a infinite plane instead of a cylinder it is also bipartite, and by $\pi/2$ phase-space rotations on all modes in one bipartion, namely every second horizontal row shown in $\textbf{Z}_\text{2D}$ above (corresponding to every second pair of modes arriving simultaneously at the homodyne detectors in Fig.~\ref{fig:setup}), we get the approximate cluster state
\begin{center}
	\includegraphics[width=0.6\textwidth]{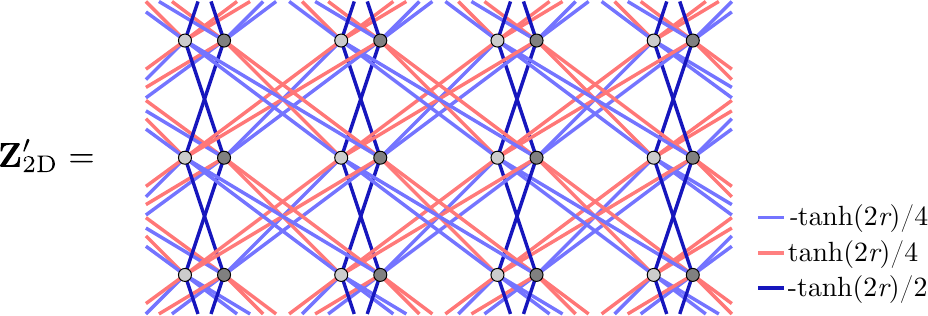},
\end{center}
where again we have omitted self-loops of $i\,\text{sech}(2r)\rightarrow 0$ for $r\rightarrow \infty$.

\begin{figure}
	\centering
	\includegraphics[width=0.5\textwidth]{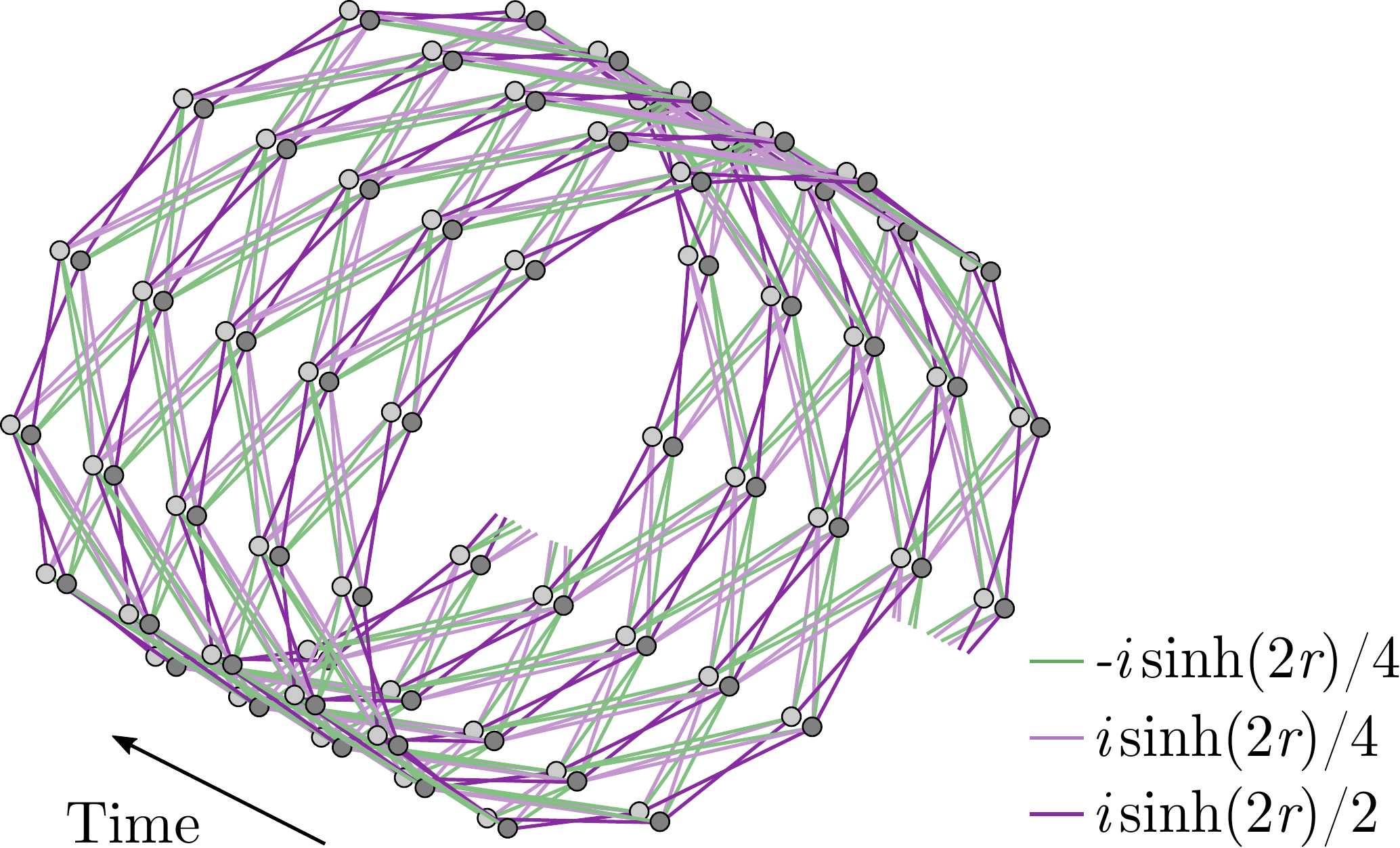}
	\caption{2D $\mathcal{H}$-graph, $\textbf{Z}_\text{2D}$, generated in Fig.~\ref{fig:setup} with $N=12$. Self-loops of $i\cosh(2r)$ are omitted for simplicity, but they are present in the diagonal of $\textbf{Z}_\text{2D}$.}
	\label{fig:2DHgrapph}
\end{figure}
Finally, considering the $\textbf{Z}_\text{2D}$ as a cylinder, the resulting $\mathcal{H}$-graph state is shown in Fig.~\ref{fig:2DHgrapph} with $N$ temporal modes in the cylinder circumference. Only in the case of even $N$, $\textbf{Z}_\text{2D}$ is a bipartite graph, and can be transformed as described above into the approximate cluster state $\textbf{Z}'_\text{2D}$ by $\pi/2$ phase-space rotation on half of its modes. As previously mentioned, such $\pi/2$ phase-space rotation of modes in the generated state can be absorbed into the measurement basis in the homodyne detection, and therefore the generated self-inverse bipartite $\mathcal{H}$-graph state is considered equivalent to its corresponding cluster state. In the experimental implementation we have chosen $N=12$ as in Fig.~\ref{fig:2DHgrapph}.

\subsection{Nullifiers}\label{sec:nullifiers}
The nullifiers of the generated 2D cluster state can be determined from its graph $\textbf{Z}'_\text{2D}$ in the same way as for the 1D cluster state in section \ref{sec:1Dcluster}. However, to give a clear picture of the quadrature transformation, here we will calculate the quadrature relations throughout the setup, from which we can finally derive the resulting nullifiers.
\begin{figure}
	\centering
	\includegraphics[width=0.6\textwidth]{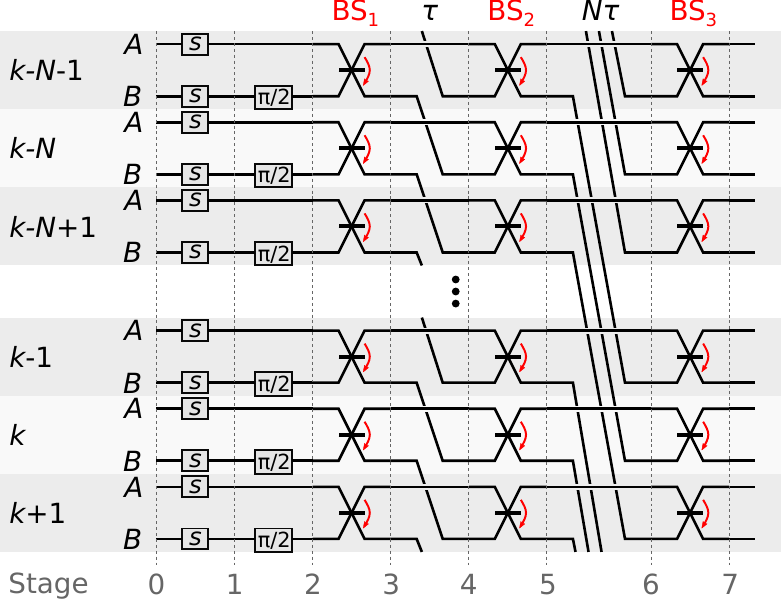}
	\caption{Corresponding circuit diagram of the experimental setup in Fig.~\ref{fig:setup} for 2D cluster state generation.}
	\label{fig:BSarray}
\end{figure}

Consider the circuit in Fig.~\ref{fig:BSarray} corresponding to the experimental setup in Fig.~\ref{fig:setup}, but with temporal modes and the effect of optical delays clearly illustrated. Here, different stages of the setup are numbered from 0 to 7, where at stage 0 all modes are initially in a vacuum state, while at stage 1 each mode are squeezed in the amplitude quadratures:
\begin{equation*}
	\hat{x}_k^{A(1)}=e^{-r_A}\hat{x}_k^{A(0)}\quad,\quad\hat{p}_k^{A(1)}=e^{r_A}\hat{p}_k^{A(0)}\quad,\quad\hat{x}_k^{B(1)}=e^{-r_B}\hat{x}_k^{B(0)}\quad,\quad\hat{p}_k^{B(1)}=e^{r_B}\hat{p}_k^{B(0)}\;,
\end{equation*}
where $r_A$ and $r_B$ are the squeezing coefficients in spatial modes $A$ and $B$ and the stage is indicated in the superscript. At stage 2, the spatial mode $B$ is rotated by $\pi/2$ in phase space such that
\begin{equation*}
	\hat{x}_k^{A(2)}=\hat{x}_k^{A(1)}=e^{-r_A}\hat{x}_k^{A(0)}\quad,\quad\hat{p}_k^{A(2)}=\hat{p}_k^{A(1)}=e^{r_A}\hat{p}_k^{A(0)}\;,
\end{equation*}
\begin{equation*}
	\hat{x}_k^{B(2)}=-\hat{p}_k^{B(1)}=-e^{r_B}\hat{p}_k^{B(0)}\quad,\quad\hat{p}_k^{B(2)}=\hat{x}_k^{B(1)}=e^{-r_B}\hat{x}_k^{B(0)}\;.
\end{equation*}
From stage 2 to 3, a beam-splitter interaction is applied onto the spatial modes $A$ to $B$,
\begin{equation*}\begin{split}
	\hat{x}_k^{A(3)}&=\frac{1}{\sqrt{2}}\left(\hat{x}_k^{A(2)}-\hat{x}_k^{B(2)}\right)=\frac{1}{\sqrt{2}}\left(e^{-r_A}\hat{x}_k^{A(0)}+e^{r_B}\hat{p}_k^{B(0)}\right)\;,\\
	\hat{p}_k^{A(3)}&=\frac{1}{\sqrt{2}}\left(\hat{p}_k^{A(2)}-\hat{p}_k^{B(2)}\right)=\frac{1}{\sqrt{2}}\left(e^{r_A}\hat{p}_k^{A(0)}-e^{-r_B}\hat{x}_k^{B(0)}\right)\;,\\
	\hat{x}_k^{B(3)}&=\frac{1}{\sqrt{2}}\left(\hat{x}_k^{A(2)}+\hat{x}_k^{B(2)}\right)=\frac{1}{\sqrt{2}}\left(e^{-r_A}\hat{x}_k^{A(0)}-e^{r_B}\hat{p}_k^{B(0)}\right)\;,\\
	\hat{p}_k^{A(3)}&=\frac{1}{\sqrt{2}}\left(\hat{p}_k^{A(2)}+\hat{p}_k^{B(2)}\right)=\frac{1}{\sqrt{2}}\left(e^{r_A}\hat{p}_k^{A(0)}+e^{-r_B}\hat{x}_k^{B(0)}\right)\;.
\end{split}\end{equation*}
From stage 3 to 4, the spatial mode $B$ is delayed by one temporal mode index,
\begin{equation*}\begin{split}
	\hat{x}_k^{A(4)}&=\hat{x}_k^{A(3)}=\frac{1}{\sqrt{2}}\left(e^{-r_A}\hat{x}_k^{A(0)}+e^{r_B}\hat{p}_k^{B(0)}\right)\;,\\
	\hat{p}_k^{A(4)}&=\hat{p}_k^{A(3)}=\frac{1}{\sqrt{2}}\left(e^{r_A}\hat{p}_k^{A(0)}-e^{-r_B}\hat{x}_k^{B(0)}\right)\;,\\
	\hat{x}_k^{B(4)}&=\hat{x}_{k-1}^{B(3)}=\frac{1}{\sqrt{2}}\left(e^{-r_A}\hat{x}_{k-1}^{A(0)}-e^{r_B}\hat{p}_{k-1}^{B(0)}\right)\;,\\
	\hat{p}_k^{B(4)}&=\hat{p}_{k-1}^{B(3)}=\frac{1}{\sqrt{2}}\left(e^{r_A}\hat{p}_{k-1}^{A(0)}+e^{-r_B}\hat{x}_{k-1}^{B(0)}\right)\;.
\end{split}\end{equation*}
From stage 4 to 5, a beam-splitter interaction is applied on the spatial modes $A$ to $B$,
\begin{equation*}\begin{split}
	\hat{x}_k^{A(5)}&=\frac{1}{\sqrt{2}}\left(\hat{x}_k^{A(4)}-\hat{x}_k^{B(4)}\right)=\frac{1}{2}\left(e^{-r_A}\left[\hat{x}_k^{A(0)}-\hat{x}_{k-1}^{A(0)}\right]+e^{r_B}\left[\hat{p}_k^{B(0)}+\hat{p}_{k-1}^{B(0)}\right]\right)\;,\\
	\hat{p}_k^{A(5)}&=\frac{1}{\sqrt{2}}\left(\hat{p}_k^{A(4)}-\hat{p}_k^{B(4)}\right)=\frac{1}{2}\left(e^{-r_B}\left[-\hat{x}_k^{B(0)}-\hat{x}_{k-1}^{B(0)}\right]+e^{r_A}\left[\hat{p}_k^{A(0)}-\hat{p}_{k-1}^{A(0)}\right]\right)\;,\\
	\hat{x}_k^{B(5)}&=\frac{1}{\sqrt{2}}\left(\hat{x}_k^{A(4)}+\hat{x}_k^{B(4)}\right)=\frac{1}{2}\left(e^{-r_A}\left[\hat{x}_{k}^{A(0)}+\hat{x}_{k-1}^{A(0)}\right]+e^{r_B}\left[\hat{p}_{k}^{B(0)}-\hat{p}_{k-1}^{B(0)}\right]\right)\;,\\
	\hat{p}_k^{B(5)}&=\frac{1}{\sqrt{2}}\left(\hat{p}_k^{A(4)}+\hat{p}_k^{B(4)}\right)=\frac{1}{2}\left(e^{-r_B}\left[-\hat{x}_{k}^{B(0)}+\hat{x}_{k-1}^{B(0)}\right]+e^{r_A}\left[\hat{p}_{k}^{A(0)}+\hat{p}_{k-1}^{A(0)}\right]\right)\;.
\end{split}\end{equation*}
From stage 5 to 6, the spatial mode $B$ is delayed by $N$ temporal modes indices,
\begin{equation*}\begin{split}
	\hat{x}_k^{A(6)}&=\hat{x}_k^{A(5)}=\frac{1}{2}\left(e^{-r_A}\left[\hat{x}_k^{A(0)}-\hat{x}_{k-1}^{A(0)}\right]+e^{r_B}\left[\hat{p}_k^{B(0)}+\hat{p}_{k-1}^{B(0)}\right]\right)\;,\\
	\hat{p}_k^{A(6)}&=\hat{p}_k^{A(5)}=\frac{1}{2}\left(e^{-r_B}\left[-\hat{x}_k^{B(0)}-\hat{x}_{k-1}^{B(0)}\right]+e^{r_A}\left[\hat{p}_k^{A(0)}-\hat{p}_{k-1}^{A(0)}\right]\right)\;,\\
	\hat{x}_k^{B(6)}&=\hat{x}_{k-N}^{B(5)}=\frac{1}{2}\left(e^{-r_A}\left[\hat{x}_{k-N}^{A(0)}+\hat{x}_{k-N-1}^{A(0)}\right]+e^{r_B}\left[\hat{p}_{k-N}^{B(0)}-\hat{p}_{k-N-1}^{B(0)}\right]\right)\;,\\
	\hat{p}_k^{B(6)}&=\hat{p}_{k-N}^{B(5)}=\frac{1}{2}\left(e^{-r_B}\left[-\hat{x}_{k-N}^{B(0)}+\hat{x}_{k-N-1}^{B(0)}\right]+e^{r_A}\left[\hat{p}_{k-N}^{A(0)}+\hat{p}_{k-N-1}^{A(0)}\right]\right)\;.
\end{split}\end{equation*}
Finally, from stage 6 to 7, a beam-splitter interaction is executed from spatial mode $A$ to $B$,
\begin{equation}\begin{aligned}\label{eq:quad}
	\hat{x}_k^{A}&=\frac{1}{\sqrt{2}}\left(\hat{x}_k^{A(6)}-\hat{x}_k^{B(6)}\right)\\
	&=\frac{1}{2\sqrt{2}}\left(e^{-r_A}\left[\hat{x}_k^{A(0)}-\hat{x}_{k-1}^{A(0)}-\hat{x}_{k-N}^{A(0)}-\hat{x}_{k-N-1}^{A(0)}\right]+e^{r_B}\left[\hat{p}_k^{B(0)}+\hat{p}_{k-1}^{B(0)}-\hat{p}_{k-N}^{B(0)}+\hat{p}_{k-N-1}^{B(0)}\right]\right)\;,\\
	\hat{p}_k^{A}&=\frac{1}{\sqrt{2}}\left(\hat{p}_k^{A(6)}-\hat{p}_k^{B(6)}\right)\\
	&=\frac{1}{2\sqrt{2}}\left(e^{-r_B}\left[-\hat{x}_k^{B(0)}-\hat{x}_{k-1}^{B(0)}+\hat{x}_{k-N}^{B(0)}-\hat{x}_{k-N-1}^{B(0)}\right]+e^{r_A}\left[\hat{p}_k^{A(0)}-\hat{p}_{k-1}^{A(0)}-\hat{p}_{k-N}^{A(0)}-\hat{p}_{k-N-1}^{A(0)}\right]\right)\;,\\
	\hat{x}_k^{B}&=\frac{1}{\sqrt{2}}\left(\hat{x}_k^{A(6)}+\hat{x}_k^{B(6)}\right)\\
	&=\frac{1}{2\sqrt{2}}\left(e^{-r_A}\left[\hat{x}_k^{A(0)}-\hat{x}_{k-1}^{A(0)}+\hat{x}_{k-N}^{A(0)}+\hat{x}_{k-N-1}^{A(0)}\right]+e^{r_B}\left[\hat{p}_k^{B(0)}+\hat{p}_{k-1}^{B(0)}+\hat{p}_{k-N}^{B(0)}-\hat{p}_{k-N-1}^{B(0)}\right]\right)\;,\\
	\hat{p}_k^{B}&=\frac{1}{\sqrt{2}}\left(\hat{p}_k^{A(6)}+\hat{p}_k^{B(6)}\right)\\
	&=\frac{1}{2\sqrt{2}}\left(e^{-r_B}\left[-\hat{x}_k^{B(0)}-\hat{x}_{k-1}^{B(0)}-\hat{x}_{k-N}^{B(0)}+\hat{x}_{k-N-1}^{B(0)}\right]+e^{r_A}\left[\hat{p}_k^{A(0)}-\hat{p}_{k-1}^{A(0)}+\hat{p}_{k-N}^{A(0)}+\hat{p}_{k-N-1}^{A(0)}\right]\right)\;,\\
\end{aligned}\end{equation}
where the superscript $(7)$ has been omitted on this final stage. Solving for the initially squeezed amplitude quadratures $e^{-r_A}\hat{x}_k^{A(0)}$ and $e^{-r_B}\hat{x}_k^{B(0)}$, a set of nullifiers are found to be
\begin{equation}\label{eq:nullX}
	\hat{n}_k^x=\hat{x}_{k}^A+\hat{x}_{k}^B-\hat{x}_{k+1}^A-\hat{x}_{k+1}^B-\hat{x}_{k+N}^A+\hat{x}_{k+N}^B-\hat{x}_{k+N+1}^A+\hat{x}_{k+N+1}^B=2\sqrt{2}e^{-r_A}\hat{x}_k^{A(0)}\;,
\end{equation}
\begin{equation}\label{eq:nullP}
	\hat{n}_k^p=\hat{p}_{k}^A+\hat{p}_{k}^B+\hat{p}_{k+1}^A+\hat{p}_{k+1}^B-\hat{p}_{k+N}^A+\hat{p}_{k+N}^B+\hat{p}_{k+N+1}^A-\hat{p}_{k+N+1}^B=-2\sqrt{2}e^{-r_B}\hat{x}_k^{B(0)}\;,
\end{equation}
with the variance
\begin{equation}\label{eq:nullVar}
	\braket{\Delta\hat{n}_k^{x2}}=4e^{-2r_A}\quad,\quad\braket{\Delta\hat{n}_k^{p2}}=4e^{-2r_B}
\end{equation}
going towards zero when $r\rightarrow\infty$ in the spatial modes $A$ and $B$. With $\hbar=1$,  $\braket{\Delta\hat{x}_k^{A(0)}}=\braket{\Delta\hat{x}_k^{B(0)}}=1/2$.

\subsection{Cluster state computation}
\noindent In the main text Fig.~{\universal}, the generated double bilayer square lattice cluster state (2xBSL), $\textbf{Z}'_\text{2D}$, is projected into a regular square lattice using mode deletion by measuring in the $\hat{x}$-basis. This serves as a proof of the generated cluster state being a universal cluster state: As the square lattice extracted from the 2xBSL is a universal resource {\citeGu}, the 2xBSL is itself a universal resource, also without projective measurements into a square lattice. In fact, one would in general optimize the projective measurements required for the specific quantum circuit to be implemented. In this section we will give an example of such optimized setting, and discuss the experimental requirements for universal quantum computation.

Modes are wasted when projecting the cluster state into a square lattice, but more important is the resulting $1/4$ edge weights of the square lattice (in the simplified picture of infinite squeezing levels). For each computation step, edge weights less than unity leads to squeezing of the state in computation as a known byproduct of the computation. In the ideal case of the cluster state prepared from momentum eigenstates, this is not a problem as the amount of byproduct squeezing is known, and can be compensated for in the following computation steps. Yet, the resource states for physical cluster state generation are finitely squeezed approximated momentum eigenstates leading to noise in the measurements of each computation step and thereby errors. These errors can be corrected by error correction (which we will come back to), thereby enabling fault-tolerant computation, as long as the squeezing of the initial resource states surpasses a given threshold. However, with the squeezing of the state in computation as byproduct in each computation step due to edge weights less than unity, less noise is required in the measurements of the computation, thereby increasing the squeezing threshold of the initial resource states for fault-tolerance. As a result, when projecting the 2xBSL into a regular square lattice with $1/4$ edge weights, not only modes are wasted, but squeezing is wasted as well.

\begin{figure}
    \centering
    \includegraphics[width=0.7\textwidth]{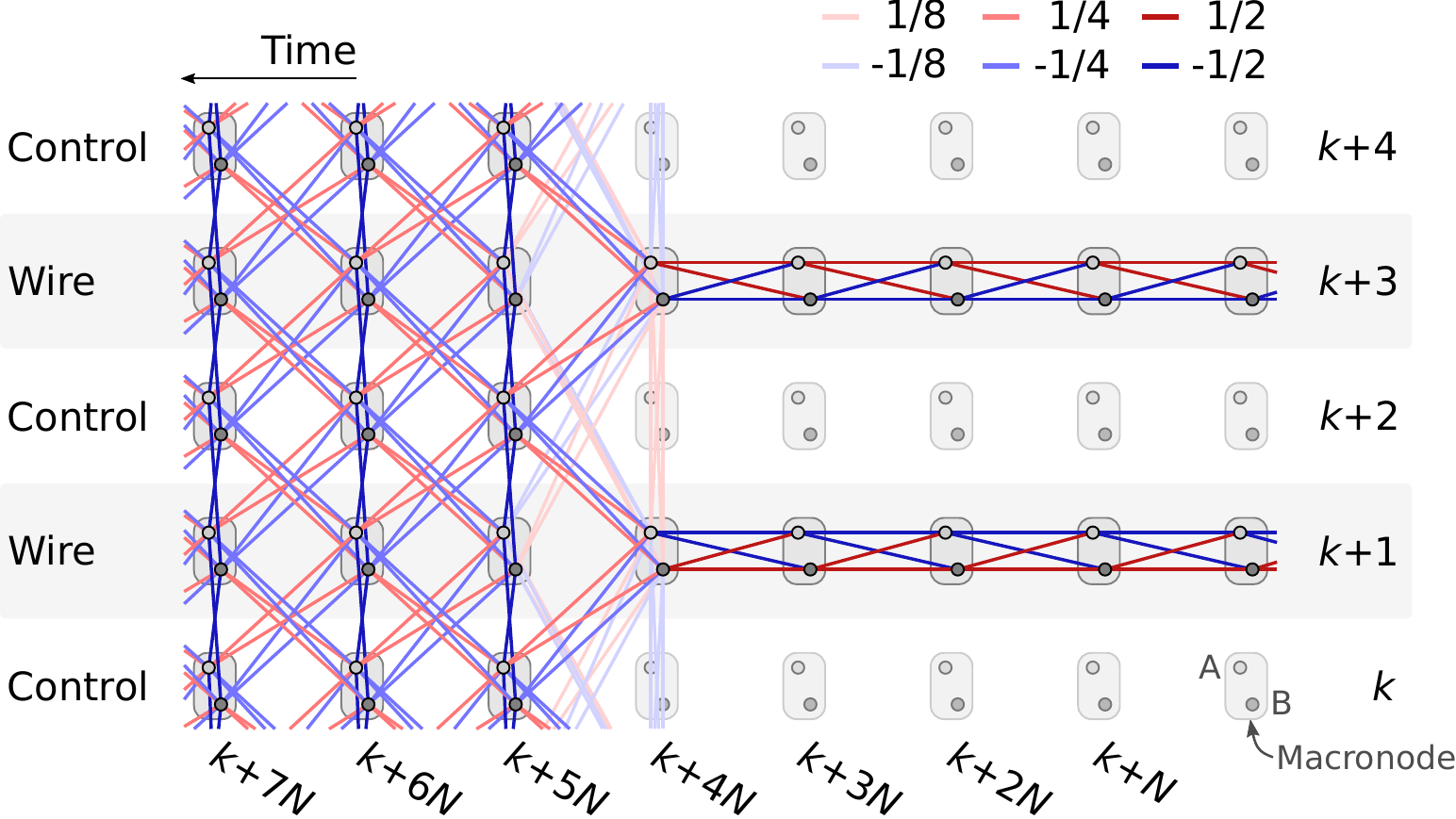}
    \caption{Depending on the measurement basis of control modes, the double bilayer square lattice can efficiently be projected into dual-rail wires with possible entanglement in between neighbouring wires depending on the basis in which the modes in the control line is measured. It is convenient to define macronodes of the two spatial modes $A$ and $B$ in the same temporal mode. In the situation depicted here, the spatial modes $A$ and $B$ of the first 5 macronodes in the control lines have been measured in bases $\left(\hat{x}\pm\hat{p}\right)/\sqrt{2}$ leading to separated dual-rail wires.}
    \label{fig:projection}
\end{figure}

To avoid squeezing waste when deleting modes by $\hat{x}$-measurements, entanglement from the modes can be rearranged in the cluster state by phase delay before measurements, or in other words, by measuring in different bases, $\hat{q}(\theta)=\hat{x}\cos\theta+\hat{p}\sin\theta$. That way, the 2xBSL can be projected into a simpler lattice without wasting squeezing. One example of this is shown in {\citeAlexander} for the bilayer square lattice cluster state (BSL), and a similar approach can be used here for the 2xBSL: Measuring the two spatial modes $A$ and $B$ for every second temporal mode $k+2n$ in basis $\hat{q}((-1)^n\pi/4)$ leads to dual-rail wires of edge weights $\pm1/2$ as illustrated in Fig.~\ref{fig:projection}. Such dual-rail wires are well known efficient resources for single mode computation, where each wire corresponds to a one-mode computer {\citeYokoyama} with input states encoded in the macronodes consisting of spatial modes $A$ and $B$ within the same temporal mode as indicated in Fig.~\ref{fig:projection}. Another benefit of projecting into these dual-rail wires is that the wires are along the length of the cylindrical structure of 2xBSL, and so it is unnecessary to ``cut up" the cylinder while the one-mode computation length can be as long as the cylinder itself. For the generated 2xBSL with $N=12$ temporal modes in the circumference, 6 parallel dual-rail wires can be constructed.

\begin{figure}
    \centering
    \includegraphics[width=0.6\textwidth]{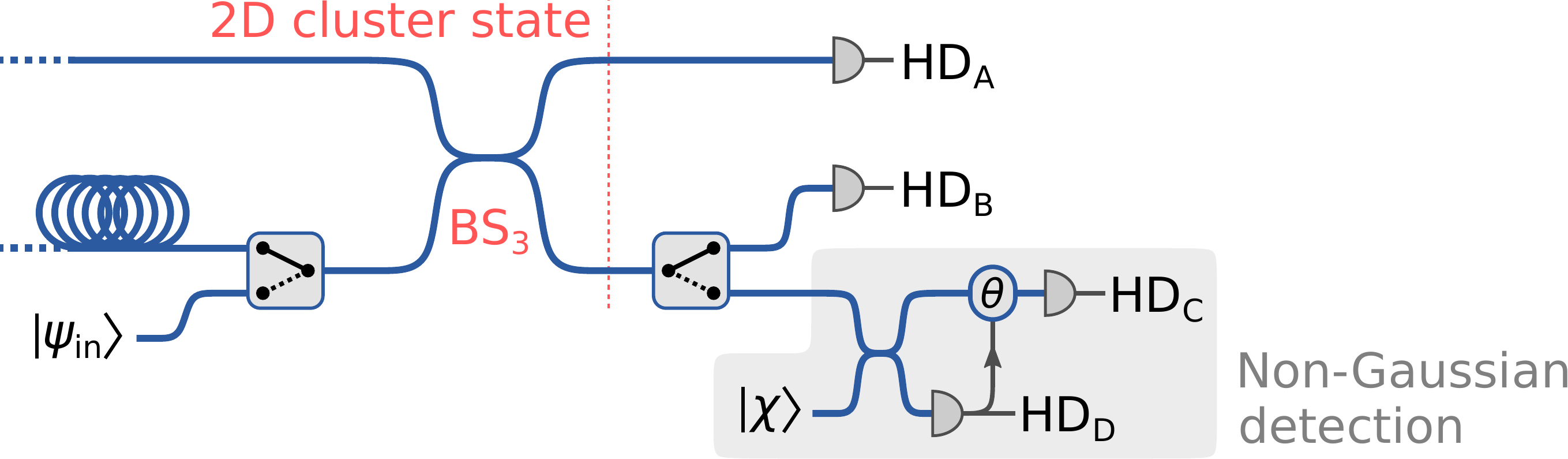}
    \caption{With a switch before beam splitter $\text{BS}_3$, input states can be encoded into the symmetric or anti-symmetric macronode given by eq.~(\ref{eq:macromode}), while Gaussian computation is performed with homodyne detectors $\text{HD}_A$ and $\text{HD}_B$. With a switch after $\text{BS}_3$ we can change to non-Gaussian measurements when required, necessary for universal quantum computation. One proposal for non-Gaussian detection is illustrated in the grey shaded area with $\ket{\chi}$ being the cubic-phase state {\citeAlexanderII}.}
    \label{fig:computation}
\end{figure}

With two spatial modes $A$ and $B$ in each temporal macronode of the dual-rail wire, the macronodes include a symmetric ($+$) and an anti-symmetric mode ($-$) via
\begin{equation}\label{eq:macromode}
    \hat{a}^\pm_k=\frac{1}{\sqrt{2}}\left(\hat{a}_k^A\pm\hat{a}_k^B\right)\;,
\end{equation}
where $\hat{a}_i$ is the annihilation operator of mode $i$, and $\hat{a}^\pm_k$ corresponds to the two spatial modes before the beam splitter $\text{BS}_3$. Thus, as in {\citeAlexander}, to encode information in a macronode, a switch can be placed before $\text{BS}_3$ as illustrated in Fig.~\ref{fig:computation}, switching the input state $\ket{\psi}_\text{in}$ into the anti-symmetric macronode with the given direction of $\text{BS}_3$ used in Fig.~\ref{fig:setup} (alternatively, we can switch into the symmetric macronode by placing the switch in spatial mode $A$ before $\text{BS}_3$). With this encoding, the multiple edges of $\pm1/2$ weight between each macronode of the dual-rail wire correspond to an edge of weight 1 between the encoded logic mode of each macronode, allowing efficient computation without squeezing waste.

Concurrently with the projection of the state into dual-rail wires, one-mode computation is performed in each wire by measuring the macronodes in an adaptive basis depending on the gate to be implemented and previous measurement outcomes. However, with the cluster state being a Gaussian state (fully described by the first and second moments of the quadratures), only Gaussian computation is possible using homodyne detectors which project Gaussian states into Gaussian states. 
For universal computation, some non-Gaussian element is needed.
As proposed in {\citeAlexanderII}, a universal gate set can be realized using the measurement scheme illustrated in the shaded area of Fig.~\ref{fig:computation}.
A switch in spatial mode $B$ will enable the choice between this measurement and the homodyne detection as needed for the computation.
With the ancillary input $\ket{\chi}$ being the highly non-Gaussian cubic-phase state, $\ket{\chi}=\int e^{i\chi s^3} \ket{s}_x \text{d}s$, this measurement implements the cubic-phase gate which, together with Gaussian gates, completes the universal one-mode gate set. Like the infinitely squeezed momentum eigenstate, the cubic-phase state is unphysical due to its infinite energy, while approximate cubic-phase states are demanding to prepare, and have not yet been generated in optical settings. Other possibilities for non-Gaussian operations exist, such as non-Gaussian projection by photon counting {\citeGottesman}, and non-Gaussian operations for a universal gate set is today an active research topic. 

For a universal multi-mode gate set, we need interaction between the dual-rail wires in Fig.~\ref{fig:projection}. By measuring macronodes between two wires in a different basis than $\hat{q}((-1)^n\pi/4)$, entanglement between the wires can be prepared depending on the measurement basis used. This is an appealing setting, where by measuring the so-called control macronodes, we can control the connectivity between neighbouring wires. However, the basis required in a macronode measurement for a given two-mode interaction is not trivial, and the detailed implementation of the $\hat{C}_Z$-gate, which together with the universal single mode gate set constitutes a universal multi-mode gate set, is left for future work.

Finally, for fault-tolerant computation error correction is necessary: With finite squeezing of the approximate momentum eigenstates from which the cluster state is generated and without error correction, only a limited number of computational steps are possible before the encoded state is lost in noise from the finite squeezing. While error correction in continuous variables is challenging, a possible way around this is to encode the information in a discrete subspace of the continuous variables concatenated with a conventional discrete error correction code. One popular example is the Gottesman-Kitaev-Preskill (GKP) encoding and correction, where a qubit is encoded as periodic peaks in the quadrature wave functions {\citeGottesman}. The squeezing level required for fault-tolerant quantum computation with the GKP-encoding depends on the error rate threshold of the concatenated error correction code to be used. For a rather conservative error correction scheme with $10^{-6}$ error rate threshold, a squeezing level of 20.5dB is required {\citeMenicucciOneFour}. With more modern error correction codes, this threshold can be brought down to 15--$\SI{17}{dB}$ of squeezing {\citeWalshe}, while the required squeezing using topological error correction is shown to be even lower {\citeFukui}.

An advantage of GKP-encoding of qubits is the recent result by B. Baragiola \textit{et al.} {\citeBaragiola} showing that the magic state can be distilled in the encoded subspace of the qubit allowing universal computation in the qubit subspace, rendering the need for non-Gaussian measurements discussed above unnecessary. However, with GKP-encoded states being highly non-Gaussian, they are as the cubic-phase state difficult to prepare, and their generation is as well an active research topic with recent demonstration in the microwave regime {\citeCampagne} and in trapped-ion mechanical oscillators {\citeFluhmann}.

\section{Inseparability criterion}\label{sec:insep}
In this section, we derive an upper bound on nullifier variance for complete inseparability of modes in the generated cluster state based on the van Loock-Furusawa criterion {\citeVanLoock}. In the van Loock-Furusawa criterion, a number of modes are divided into two or more sets from which an inequality with combined quadrature variance is derived.
A violation of this inequality means that the sets are inseparable.

For simplicity, we will consider only two sets of modes, $\mathcal{S}_1$ and $\mathcal{S}_2$, and define
\begin{equation}\label{eq:QP}
	\hat{X}=\sum_{j\in \mathcal{S}_1\cup \mathcal{S}_2}h_j \hat{x}_j\quad,\quad \hat{P}=\sum_{j\in \mathcal{S}_1\cup \mathcal{S}_2}g_j \hat{p}_j\;,
\end{equation}
for arbitrary coefficients $h_j$ and $g_j$. The van Loock-Furusawa criterion for separability then reads
\begin{equation}\label{eq:LFc}
	\braket{\Delta \hat{X}^2}+\braket{\Delta \hat{P}^2} \geq \Big| \sum_{j\in \mathcal{S}_1} h_jg_j\Big|+\Big| \sum_{j\in \mathcal{S}_2} h_jg_j\Big|\;,
\end{equation}
with $\hbar=1$. The goal is to find suitable $h_j$ and $g_j$ such that eq.~(\ref{eq:LFc}) is violated, thus proving inseparability of the two sets. Doing so for all possible bipartitions of modes then proves complete inseparability.

Since the generated cluster state is periodic, it is only necessary to consider the modes of a single unit cell of the cluster state lattice, and show complete inseparability of the modes within this unit cell. A good example of this approach is shown in the supplementary material of {\citeYoshikawa} for a 1D cluster states. The 8 modes of the nullifiers $\hat{n}_k^x$ and $\hat{n}_k^p$ in eq.~(\ref{eq:nullX}) and (\ref{eq:nullP}) make up a unit cell of the generated 2D cluster state, and is illustrated in Fig.~\ref{fig:inseparability} with the modes numbered from 1 to 8. 
Hence, complete inseparability of the 2D cluster state can be proven by demonstrating a violation of the separability inequality in eq.~(\ref{eq:LFc}) for each of the $2^{8-1}-1=127$ possible bipartitions of these 8 modes. Below, we give three examples with different bipartitions. The mode numbering in Fig.~\ref{fig:inseparability} is used to shorten the notation:
\begin{figure}
	\centering
	\includegraphics[width=0.4\textwidth]{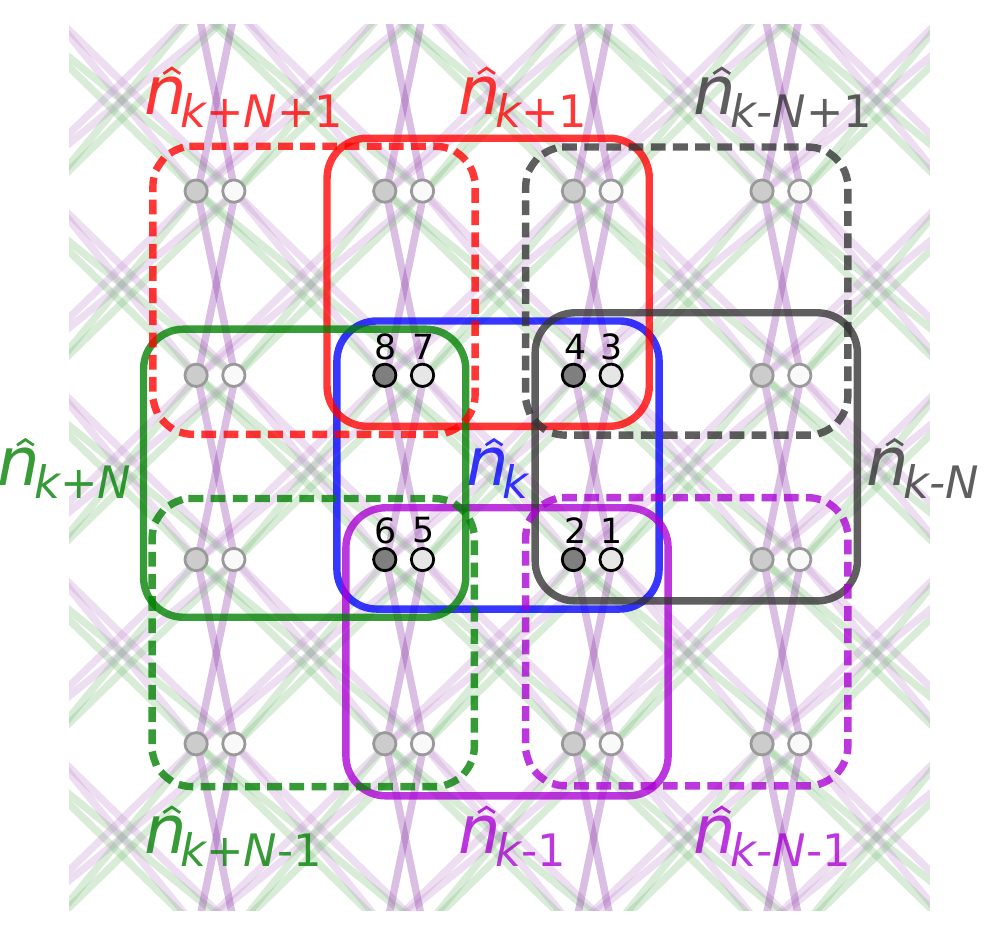}
	\caption{Graph of the generated 2D cluster state with the nullifier $\hat{n}_k$ ($\hat{n}_k^x$ or $\hat{n}_k^p$) and its neighbouring nullifiers indicated. In the van Loock-Furusawa inseparability criterion we consider a unit cell of 8 modes in common with $\hat{n}_k$, numbered as $(A,k)\rightarrow1$, $(B,k)\rightarrow2$, $(A,k+1)\rightarrow3$, $(B,k+1)\rightarrow4$, $(A,k+N)\rightarrow5$, $(B,k+N)\rightarrow6$, $(A,k+N+1)\rightarrow7$ and $(B,k+N+1)\rightarrow8$.}
	\label{fig:inseparability}
\end{figure}
\\
\\
\textbf{Example 1:} Consider the two sets of modes $\mathcal{S}_1=\{1,2,5,6\}$ and $\mathcal{S}_2=\{3,4,7,8\}$. Choosing
\begin{equation*}
	\hat{X}=\hat{n}_k^x= \hat{x}_1 + \hat{x}_2 - \hat{x}_3 - \hat{x}_4 - \hat{x}_5 + \hat{x}_6 - \hat{x}_7 + \hat{x}_8\
\end{equation*}
such that $(h_1,h_2,h_3,h_4,h_5,h_6,h_7,h_8)=(1,1,-1,-1,-1,1,-1,1)$, and 
\begin{equation*}
	\hat{P}=\hat{n}_k^p = \hat{p}_1 + \hat{p}_2 + \hat{p}_3 + \hat{p}_4 - \hat{p}_5 + \hat{p}_6 + \hat{p}_7 - \hat{p}_8
\end{equation*}
such that $(g_1,g_2,g_3,g_4,g_5,g_6,g_7,g_8)=(1,1,1,1,-1,1,1-1)$, then eq.~(\ref{eq:LFc}) becomes
\begin{equation*}\begin{split}
	\braket{\Delta \hat{X}^2}+\braket{\Delta \hat{P}^2}=\braket{\Delta n_k^x{}^2}+\braket{\Delta n_k^p{}^2}\geq&\Big| \sum_{j\in S_1} h_jg_j\Big|+\Big| \sum_{j\in S_2} h_jg_j\Big|\\
	 &= |1\cdot1+ 1\cdot1+(-1)\cdot(-1)+1\cdot1|\\
	 &\quad\quad+|(-1)\cdot1+(-1)\cdot1+(-1)\cdot1+1\cdot(-1)|\\
	 &=8\;.
\end{split}\end{equation*}
We may measure different variances of $\hat{n}_k^x$ and $\hat{n}_k^p$, but if we measure both below 4, the above inequality will for sure be violated and the two mode sets $\mathcal{S}_1$ and $\mathcal{S}_2$ are inseparable. From eq.~(\ref{eq:nullVar}) this requires $4e^{-2r_i}<4$ for $i=A,B$, and thus measuring the variance of $\hat{n}_k^x$ and $\hat{n}_k^p$ with more than $\SI{0}{dB}$ squeezing below shot noise.\\
\\
\textbf{Example 2:} Consider now the two mode sets $\mathcal{S}_1=\{1,2,3,4\}$ and $\mathcal{S}_2=\{5,6,7,8\}$. Choosing $\hat{X}$ and $\hat{P}$ as in example 1 leads to
\begin{equation*}\begin{split}
	\braket{\Delta \hat{X}^2}+\braket{\Delta \hat{P}^2}=\braket{\Delta \hat{n}_k^x{}^2}+\braket{\Delta \hat{n}_k^p{}^2}\geq&| \sum_{j\in S_1} h_jg_j|+| \sum_{j\in S_2} h_jg_j|\\
	 &= |1\cdot1+ 1\cdot1+(-1)\cdot1+(-1)\cdot1|\\
	 &\quad\quad+|(-1)\cdot(-1)+1\cdot1+(-1)\cdot1+1\cdot(-1)|\\
	 &=0\;,
\end{split}\end{equation*}
which is impossible to violate. If we instead choose
\begin{equation*}
	\hat{P}=\hat{n}_{k+1}^p=\hat{p}_3+\hat{p}_4-\hat{p}_7+\hat{p}_8+\hat{p}_{k+2}^A+\hat{p}_{k+2}^B+\hat{p}_{k+N+2}^A+\hat{p}_{k+N+2}^B\;,
\end{equation*}
such that $(g_1,g_2,g_3,g_4,g_5,g_6,g_7,g_8)=(0,0,1,1,0,0,-1,1)$, eq.~(\ref{eq:LFc}) becomes
\begin{equation}\begin{split}\label{eq:ex2}
	\braket{\Delta \hat{X}^2}+\braket{\Delta \hat{P}^2}=\braket{\Delta \hat{n}_k^x{}^2}+\braket{\Delta \hat{n}_{k+1}^{p2}}\geq&| \sum_{j\in S_1} h_jg_j|+| \sum_{j\in S_2} h_jg_j|\\
	 &= |1\cdot0+ 1\cdot0+(-1)\cdot1+(-1)\cdot1|\\
	 &\quad\quad+|(-1)\cdot0+1\cdot0+(-1)\cdot(-1)+1\cdot1|\\
	 &=4\;,
\end{split}\end{equation}
which is violated if the variance of the two nullifiers $\hat{n}_k^x$ and $\hat{n}_k^p$ are less than 2, requiring $\SI{3}{dB}$ of squeezing. The additional modes included in $\hat{P}$, $(A,k+2)$, $(B,k+2)$, $(A,k+N+2)$ and $(B,k+N+2)$, are not included in the above inequality since they are not common modes with any in $\hat{X}=\hat{n}_k^x$, and thus will not contribute to the right hand side of eq.~(\ref{eq:LFc}). However, when including 4 extra modes, we should consider all new possible bipartitions: Given the two sets $\mathcal{S}_1$ and $\mathcal{S}_2$, we can add the additional 4 modes into these two sets in any arbitrary way without any change to  eq.~(\ref{eq:LFc}). As a result, by violating eq.~(\ref{eq:LFc}) we prove inseparability of all bipartitions where each of the 4 extra modes are added to $\mathcal{S}_1$ or $\mathcal{S}_2$ in all possible ways.\\
\\
\textbf{Example 3:} Consider the two mode sets $\mathcal{S}_1=\{3,5\}$ and $\mathcal{S}_2=\{1,2,4,6,7,8\}$. For this bipartition, there exists no single nullifier for $\hat{X}$ and $\hat{P}$ of the form in eq.~(\ref{eq:nullX}) and (\ref{eq:nullP}) which forms an inequality we can hope to violate experimentally. However, since linear combinations of nullifiers are also nullifiers, more exotic choices for $\hat{X}$ and $\hat{P}$ exist which leads to an inequality we can violate experimentally:
\begin{equation*}\begin{split}
	\hat{X}&=-\hat{n}_k^x+\hat{n}_{k+N}^x=-\hat{x}_1-\hat{x}_2+\hat{x}_3+\hat{x}_4+2\hat{x}_5-2\hat{x}_8-\hat{x}_{k+2N}^A+\hat{x}_{k+2N}^B-\hat{x}_{k+2N+1}^A+\hat{x}_{k+2N+1}^B\\
	\\
	\hat{P}&=\hat{n}_{k-1}^p+\hat{n}_{k+N}^p+\hat{n}_{k+N+1}^p+\hat{n}_{k+N-1}^p\\
	&=\hat{p}_1+\hat{p}_2+3\hat{p}_5+\hat{p}_6+2\hat{p}_7+2\hat{p}_8+\hat{p}_{k-1}^A+\hat{p}_{k-1}^B+2\hat{p}_{k+N-1}^B\\
	&\hspace{4cm}+\hat{p}_{k+N+2}^A+\hat{p}_{k+N+2}^B-\hat{p}_{k+2N-1}^A+\hat{p}_{k+2N-1}^B+\hat{p}_{k+2N+2}^A-\hat{p}_{k+2N+2}^B
\end{split}\end{equation*}
leading to 
\begin{equation*}
(h_1,h_2,h_3,h_4,h_5,h_6,h_7,h_8)=(-1,-1,1,1,2,0,0,-2)\;,\;(g_1,g_2,g_3,g_4,g_5,g_6,g_7,g_8)=(1,1,0,0,3,1,2,2)\;,
\end{equation*}
and so (\ref{eq:LFc}) becomes
\begin{equation*}\begin{split}
	\braket{\Delta \hat{X}^2}+\braket{\Delta \hat{P}^2}=&\braket{\Delta \hat{n}_k^{x2}}+\braket{\Delta \hat{n}_{k+N}^{x2}}+\braket{\Delta \hat{n}_{k-1}^{p2}}+\braket{\Delta \hat{n}_{k+N}^{p2}}+\braket{\Delta \hat{n}_{k+N+1}^{p2}}+\braket{\Delta \hat{n}_{k+N-1}^{p2}}\\
	\geq&| \sum_{j\in S_1} h_jg_j|+| \sum_{j\in S_2} h_jg_j|\\
	 =& |1\cdot0+ 2\cdot3|+|(-1)\cdot1+(-1)\cdot1+1\cdot0+0\cdot1+0\cdot2+(-2)\cdot 2|=12\;,
\end{split}\end{equation*}
which is violated if the variance of each of the 6 nullifiers in the inequality is less than 2, corresponding to $\SI{3}{dB}$ of squeezing. 
Here, the cross terms between different nullifiers in $\braket{\Delta\hat{X}^2}$ and $\braket{\Delta\hat{P}^2}$ are zero, as the nullifiers in eqs. (\ref{eq:nullX}) and (\ref{eq:nullP}) correspond to different temporal modes at the squeezing sources before $\text{BS}_1$, and thus there are no correlations between different nullifiers.
Notice how the nullifiers in $\hat{X}$ and $\hat{P}$ are chosen such that the 4 and 9 extra modes included in $\hat{X}$ and $\hat{P}$, respectively, are not the same. Thus, by the same argument as in example 2, violating the above inequality proves inseparability of all bipartitions where each of the $4+9$ extra modes are added to $\mathcal{S}_1$ or $\mathcal{S}_2$ in all possible ways.\\

Using the same approach as in the above three examples, nullifiers for $\hat{X}$ and $\hat{P}$ are found for all 127 possible bipartitions of the 8 modes in the studied unit cell, resulting in a sufficient condition for the squeezing degree among all nullifiers of $\SI{3}{dB}$ below shot noise. As a result, with the generated 2D cluster state being periodic with this unit cell, measuring every temporal nullifier ($\hat{n}_k^x$ and $\hat{n}_k^p$ for every $k$) with a variance less than $\SI{3}{dB}$ below shot noise leads to complete inseparability of the cluster state. The resulting $\hat{X}$ and $\hat{P}$ for every bipartition are listed in Table S1 in the end of this supplementary material, and as pointed out in example 2 and 3, each choice of $\hat{X}$ and $\hat{P}$ are made such that they do not share any modes outside the studied unit cell.

\section{Experimental setup}\label{sec:expMethods}
In this section, more details and characterization is given on the experimental implementation described in Materials and Methods section in the beginning of this supplementary materials.

\subsection{Efficiency and phase stability}\label{sec:eff_and_phase}
In the following, all loss contributions are summarized, and a combined efficiency of the setup is estimated: The $\text{OPO}_\text{A}$ and $\text{OPO}_\text{B}$ escape efficiencies are measured to be 0.98 and 0.95, respectively, while 1\% is tapped off in both squeezing sources for gain locks. The two spatial modes, $A$ and $B$, are coupled from free-space into fiber with a 0.97 coupling efficiency, where $3\times1\%$ is tapped off for phase locking the three interference points at $\text{BS}_1$, $\text{BS}_2$ and $\text{BS}_3$, each with an estimated visibility of 0.99. To minimize the propagation losses, all fibers are spliced together, while short and long delay lines of SMF-28e+ fiber with $\SI{0.2}{dB/km}$ attenuation each leads to $0.2\%$ and $2.7\%$ propagation loss, respectively. Finally, the fiber based homodyne detectors each have a detection efficiency of 0.91. For more information on the OPO, fiber coupling and homodyne detection efficiencies, see {\citeLarsen}. In total, the estimated efficiencies add up to 0.81 and 0.78 in spatial mode $A$ and $B$, respectively.

\begin{figure}
	\centering
	\includegraphics[width=0.6\textwidth]{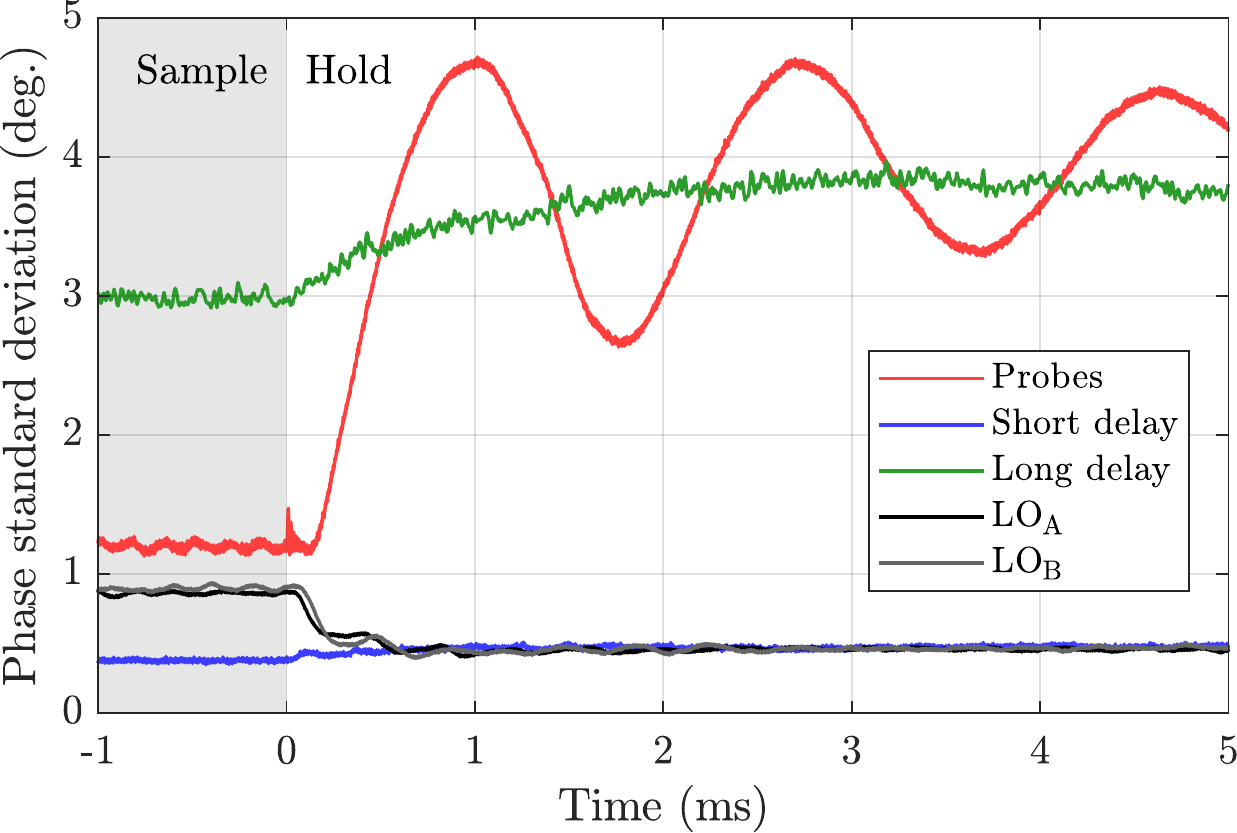}
	\caption{Standard deviation of the phase fluctuations in different parts of the setup measured by sending a coherent probe through the particular part of the setup while feedback is kept constant (hold-time). The phase fluctuations of the short and long delay lengths are measured by coupling a probe into the setup before $\text{BS}_1$ and $\text{BS}_2$ while measuring the interference after $\text{BS}_2$ and $\text{BS}_3$ respectively. The relative phase fluctuation of the two probes from $\text{OPO}_\text{A}$ and $\text{OPO}_\text{B}$ is measured by the interference after $\text{BS}_1$. The local oscillator (LO) phase fluctuations are measured by coupling a probe into the setup before $\text{BS}_3$ and measuring the probe quadrature.}
	\label{fig:phase}
\end{figure}
Besides loss, the generated 2D cluster state is affected by phase fluctuations. In Fig.~\ref{fig:phase}, the standard deviation of the phase is shown. The phases were measured while probing different parts of the setup with a coherent beam while turning off the feedback for cavity or phase locks. As expected, we see around 6 times more phase fluctuation of the long delay line compared to the short delay line. Another, and maybe more surprising, contribution to the phase fluctuation is from the probe phase which is seen to fluctuate fast as soon as the feedback is kept constant (hold-time). This is explained by the strong phase dependence in the OPO cavities around resonance. Furthermore, the probe phase standard deviation is seen to fluctuate, indicating systematic phase fluctuations which we believe are due to mechanical resonance and limited feedback bandwidth leading to a large impulse response when the feedback is suddenly kept constant when changing from sample- to hold-time. However, from this phase measurement, it is not clear whether the large phase fluctuation is from the probes of both OPO cavities, or if mainly one OPO cavity is more unstable. Finally, the standard deviation of the local oscillator (LO) phases appears to decrease during hold-time. This is simply caused by the fact that the probe quadrature fluctations (in addition to the LO noise) are measured during the sample-time while during the hold-time, only the LO noise is measured.

\subsection{Spectrum}\label{sec:spectrum}
\begin{figure}
	\centering
	\includegraphics[width=0.9\textwidth]{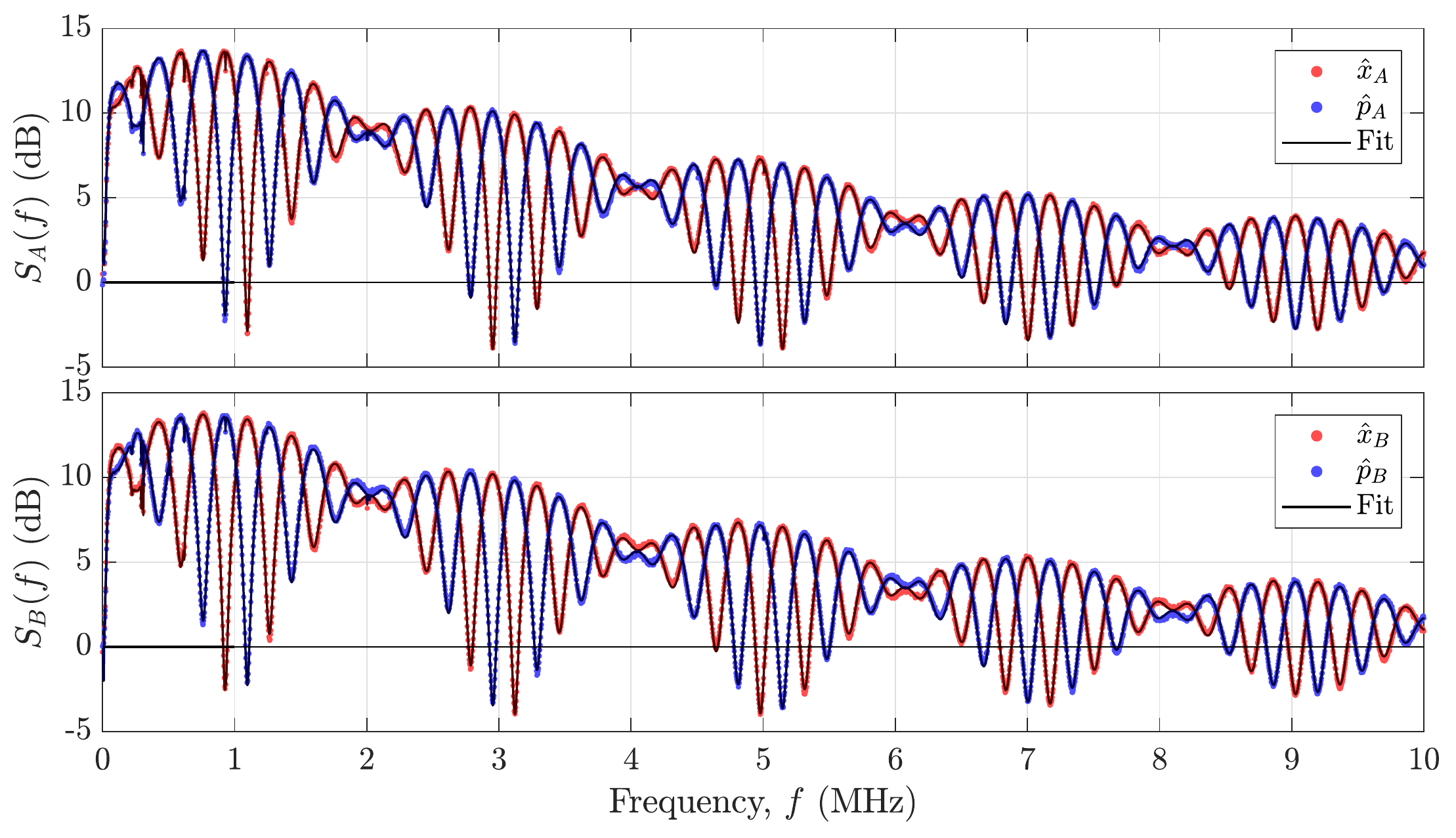}
	\caption{Power spectrum of the temporal encoded cluster state measured at the homodyne detectors in the spatial mode $A$ and $B$. The solid lines shows the result of fitting the power spectra in eq.~(\ref{eq:Spectra}) with the squeezing spectra in eq.~(\ref{eq:SqSpectra}) including phase fluctuations by eq.~(\ref{eq:phase}) and measured electronic noise. The resulting fitting parameters are listed in eq.~(\ref{eq:fitResult}).}
	\label{fig:spectrum}
\end{figure}

The generated 2D cluster state is temporally encoded in 2 spatial modes, $A$ and $B$. As a result, modes of the cluster state are measured by acquiring time traces from the two homodyne detectors in $A$ and $B$, on which a temporal mode function is applied for each mode as will be described in section \ref{sec:TMF}. However, by analyzing the acquired time traces in frequency domain, we can obtain useful information about the setup and the two squeezing sources. In Fig.~\ref{fig:spectrum} the power spectra of the acquired time traces are shown, calculated by fast Fourier transform of $\SI{320}{\micro s}$ long time traces corresponding to $\SI{1300}{}$ consecutive temporal modes. To understand these power spectra, we derive them theoretically in the following.

According to the Wiener-Khinchin theorem, the quadrature power spectrum is expressed by the Fourier transform of the quadrature autocorrelation function,
\begin{equation}\label{eq:S}
	S_j^q(\omega)=\int_{-\infty}^\infty\braket{\hat{q}_j(t)\hat{q}_j(0)}e^{i\omega t}\,\text{d}t\quad,\quad j=A,B\;,\;q=x,p\;,
\end{equation}
where $\omega$ is the angular frequency. The time dependent amplitude and phase quadratures, $\hat{x}(t)$ and $\hat{p}(t)$, are derived in the exact same way as the temporal mode quadratures in eq.~(\ref{eq:quad}), and the result is the same but with time dependency instead of temporal mode index, i.e. $\hat{q}_{j,k-m}\rightarrow\hat{q}_j(t-m\tau)$ with $m=0,1,N,N+1$, as neighbouring temporal modes are spaced in time by $\tau$. Considering first the $\hat{x}$-quadrature in mode $A$, the autocorrelation function, using the quadratures expressed in eq.~(\ref{eq:quad}), becomes
\begin{equation*}\begin{split}
	\braket{\hat{x}_A(t)\hat{x}_A(0)}=&\frac{1}{8}\left(4\braket{\hat{x}_A^{(1)}(t)\hat{x}_A^{(1)}(0)}-\braket{\hat{x}_A^{(1)}(t+N\tau+\tau)\hat{x}_A^{(1)}(0)}+\braket{\hat{x}_A^{(1)}(t+N\tau-\tau)\hat{x}_A^{(1)}(0)}\right.\\
	&\hspace{4cm}\left.+\braket{\hat{x}_A^{(1)}(t-N\tau+\tau)\hat{x}_A^{(1)}(0)}-\braket{\hat{x}_A^{(1)}(t-N\tau-\tau)\hat{x}_A^{(1)}(0)}\right)\\
	&+\frac{1}{8}\left(4\braket{\hat{p}_B^{(1)}(t)\hat{p}_B^{(1)}(0)}+\braket{\hat{p}_B^{(1)}(t+N\tau+\tau)\hat{p}_B^{(1)}(0)}-\braket{\hat{p}_B^{(1)}(t+N\tau-\tau)\hat{p}_B^{(1)}(0)}\right.\\
	&\hspace{4cm}\left.-\braket{\hat{p}_B^{(1)}(t-N\tau+\tau)\hat{p}_B^{(1)}(0)}+\braket{\hat{p}_B^{(1)}(t-N\tau-\tau)\hat{p}_B^{(1)}(0)}\right)\;,
\end{split}\end{equation*}
where the property of the autocorrelation $\braket{\hat{q}_j(t)\hat{q}_j(y)}=\braket{\hat{q}_j(t-y)\hat{q}_j(0)}$ is used. Substituting $\braket{\hat{x}_A(t)\hat{x}_A(0)}$ into eq.~(\ref{eq:S}), and using that
\begin{equation*}
	\int_{-\infty}^\infty\braket{\hat{q}_j(t+y)\hat{q}_j(0)}e^{i\omega t}\,\text{d}t=	\int_{-\infty}^\infty\braket{\hat{q}_j(t)\hat{q}_j(0)}e^{i\omega (t-y)}\,\text{d}t=e^{-i\omega y}S_j^{q}(\omega)\;,
\end{equation*}
the power spectrum measured in mode $A$ in the $\hat{x}$-quadrature becomes
\begin{equation*}\begin{split}
	S_A^x(\omega)=&\frac{1}{8}\left(4-e^{-i\omega(N\tau+\tau)}+e^{-i\omega(N\tau-\tau)}+e^{-i\omega(-N\tau+\tau)}-e^{-i\omega(-N\tau-\tau)}\right)S_A^{x(1)}(\omega)\\
	&\hspace{1cm}+\frac{1}{8}\left(4+e^{-i\omega(N\tau+\tau)}-e^{-i\omega(N\tau-\tau)}-e^{-i\omega(-N\tau+\tau)}+e^{-i\omega(-N\tau-\tau)}\right)S_B^{p(1)}(\omega)\\
	=&\frac{1}{4}\big(2-\cos(\omega N\tau+\omega\tau)+\cos(\omega N\tau-\omega\tau)\big)S_A^{x(1)}(\omega)\\
	&\hspace{1cm}+\frac{1}{4}\big(2+\cos(\omega N\tau+\omega\tau)-\cos(\omega N\tau-\omega\tau)\big)S_B^{p(1)}(\omega)\;,
\end{split}\end{equation*}
where $S_A^{x(1)}$ and $S_B^{p(1)}$ are power spectra at stage 1 in Fig.~\ref{fig:BSarray}, and corresponds to the squeezing and the anti-squeezing spectrum of the amplitude squeezing sources in mode $A$ and $B$ respectively. Following the same approach for the $\hat{p}$-quadrature and for the quadratures in mode $B$, and using that $2\mp\cos(\omega N\tau+\omega\tau)\pm\cos(\omega N\tau-\omega\tau)=2\pm2\sin(\omega N\tau)\sin(\omega\tau)$, the power spectra displayed in Fig.~\ref{fig:spectrum} are expressed as
\begin{equation}\begin{split}\label{eq:Spectra}
	&S_A^x(\omega)= \frac{1}{2}\big(1+\sin(\omega N\tau)\sin(\omega\tau)\big)S_A^{x(1)}(\omega)+\frac{1}{2}\big(1-\sin(\omega N\tau)\sin(\omega\tau)\big)S_B^{p(1)}(\omega)\;,\\
	&S_A^p(\omega)= \frac{1}{2}\big(1-\sin(\omega N\tau)\sin(\omega\tau)\big)S_B^{x(1)}(\omega)+\frac{1}{2}\big(1+\sin(\omega N\tau)\sin(\omega\tau)\big)S_A^{p(1)}(\omega)\;,\\
	&S_B^x(\omega)= \frac{1}{2}\big(1-\sin(\omega N\tau)\sin(\omega\tau)\big)S_A^{x(1)}(\omega)+\frac{1}{2}\big(1+\sin(\omega N\tau)\sin(\omega\tau)\big)S_B^{p(1)}(\omega)\;,\\
	&S_B^p(\omega)= \frac{1}{2}\big(1+\sin(\omega N\tau)\sin(\omega\tau)\big)S_B^{x(1)}(\omega)+\frac{1}{2}\big(1-\sin(\omega N\tau)\sin(\omega\tau)\big)S_A^{p(1)}(\omega)\;.
\end{split}\end{equation}
Finally, the squeezing spectra $S_j^{q(1)}$ from the OPO squeezing sources, squeezed in the amplitude quadrature, are derived in {\citeCollett} to be
\begin{equation}\label{eq:SqSpectra}
	S_j^{x(1)}(\omega)=\frac{1}{2}-\frac{2\varepsilon_j\gamma_j\eta_j}{(\gamma_j+\varepsilon_j)^2+\omega^2}\quad,\quad S_j^{p(1)}(\omega)=\frac{1}{2}+\frac{2\varepsilon_j\gamma_j\eta_j}{(\gamma_j-\varepsilon_j)^2+\omega^2}\quad, \;j=A,B\;,
\end{equation}
where $\varepsilon_j$, $\gamma_j$ and $\eta_j$ is the pump rate, total OPO decay rate and squeezing source efficiency, respectively, in mode $j=A,B$.

To include phase fluctuations in the spectra, we should ideally include phase fluctuation in the quadrature transformation at every stage in Fig.~\ref{fig:BSarray}. However, for simplicity, we include all phase fluctuations either before or after the beam-splitter array from stage 2 to 7. Since the sensitive OPO cavities are one of the dominating sources of phase fluctuations, here we include phase fluctuations in the squeezing source, i.e. at stage 1. Assuming the statistics of the phase fluctuations to follow a normal distribution of phase, $\theta$, with the width $\sigma$, $P(\theta,\sigma)$, the phase fluctuations are included in the squeezing spectrum as
\begin{equation}\begin{aligned}\label{eq:phase}
	S_j^{x(1)}(\omega,\sigma)=&\int P(\sigma_j,\theta)\left(S_j^{x(1)}(\omega)\cos^2\theta+S_j^{p(1)}\sin^2\theta\right)\,\text{d}\theta&\\
	\approx& S_j^{x(1)}(\omega)\cos^2\sigma_j+S_j^{p(1)}\sin^2\sigma_j&,\;j=A,B\;,
\end{aligned}\end{equation}
where the approximation holds for small $\sigma$, and the same for $S_j^{p(1)}(\omega,\sigma)$ with $\cos$ and $\sin$ interchanged.

In Fig.~\ref{fig:spectrum}, we present the fitted power spectra of eq.~(\ref{eq:Spectra}) accounting for phase fluctuations (as in eq.~(\ref{eq:phase})) and electronic noise by including a frequency dependent electronic efficiency determined from a measured electronic power spectrum. The fitting parameters are $\varepsilon_j$, $\gamma_j$, $\eta_j$ and $\sigma_j$ ($j=A,B$) and we use $N=12$ and $\SI{247}{ns}$. The result of the fitting routine is
\begin{equation}\begin{aligned}\label{eq:fitResult}
	\varepsilon_A&=2\pi\times 5.38\pm\SI{0.02}{MHz}\hspace{1cm}&\varepsilon_B&=2\pi\times5.57\pm\SI{0.02}{MHz}\\
	\gamma_A&=2\pi\times7.59\pm\SI{0.02}{MHz}\quad&\gamma_B&=2\pi\times7.80\pm\SI{0.02}{MHz}\\
	\eta_A&=0.789\pm0.004\quad&\eta_B&=0.764\pm0.004\\
	\sigma_A&=5.17\pm0.12^\circ\quad&\sigma_B&=5.90\pm0.10^\circ\;,
\end{aligned}\end{equation}
where uncertainties are estimated as the 95\% confidence interval. The fit is seen to agree very well with the measured data, and supports $N=12$ with $\tau=\SI{247}{ns}$. The fitted $\eta_A$ and $\eta_B$ differ by 0.025, which is expected due to 3\% lower escape efficiency of the $\text{OPO}_\text{B}$ compared to $\text{OPO}_\text{A}$. The fitted OPO decay rates are as expected for the OPO design, while $\text{OPO}_\text{B}$ is pumped slightly harder to compensate for the lower escape efficiency. Both OPOs are pumped to around half the threshold ($\varepsilon^2/\gamma^2=0.50$ for $\text{OPO}_\text{A}$ and $0.51$ for $\text{OPO}_\text{B}$). The fitted phase fluctuations, $\sigma_A$ and $\sigma_B$, are seen to be comparable with the measured phase fluctuations in Fig.~\ref{fig:phase}. However, with the model used for the phase fluctuations, $\sigma_A$ and $\sigma_B$ do not represent the phase fluctuation of the squeezing sources only, but a combination of phase fluctuations throughout the setup, and thus we cannot conclude the squeezing sources to have similar phase fluctuation from this fit. Finally, $\eta_A$ and $\eta_B$ are not only the efficiency of the squeezing sources, but includes efficiency throughout the setup, and can be compared with the estimated efficiencies in section \ref{sec:eff_and_phase} of 0.81 and 0.78 in spatial mode $A$ and $B$ respectively. The fitted efficiency is slightly lower than the estimated efficiency, which may be explained by experimental imperfections (e.g. lossy fiber splicing and polarization drift) which are not included in the estimation.

\subsection{Temporal mode function}\label{sec:TMF}
A temporal mode $k$ is defined by its temporal mode function $f_k(t)$. In the experimental setup a quadrature, $\hat{q}(t)$, is continuously measured by homodyne detection, and by integrating the acquired quadrature time trace weighted by the temporal mode function, we obtain the measured quadrature of the corresponding temporal mode,
\begin{equation*}
	\hat{q}_k=\int f_k(t)\hat{q}(t)\,\text{d}t\;.
\end{equation*}

\begin{figure}
	\centering
	\begin{subfigure}{0.48\textwidth}
		\includegraphics[width=\textwidth]{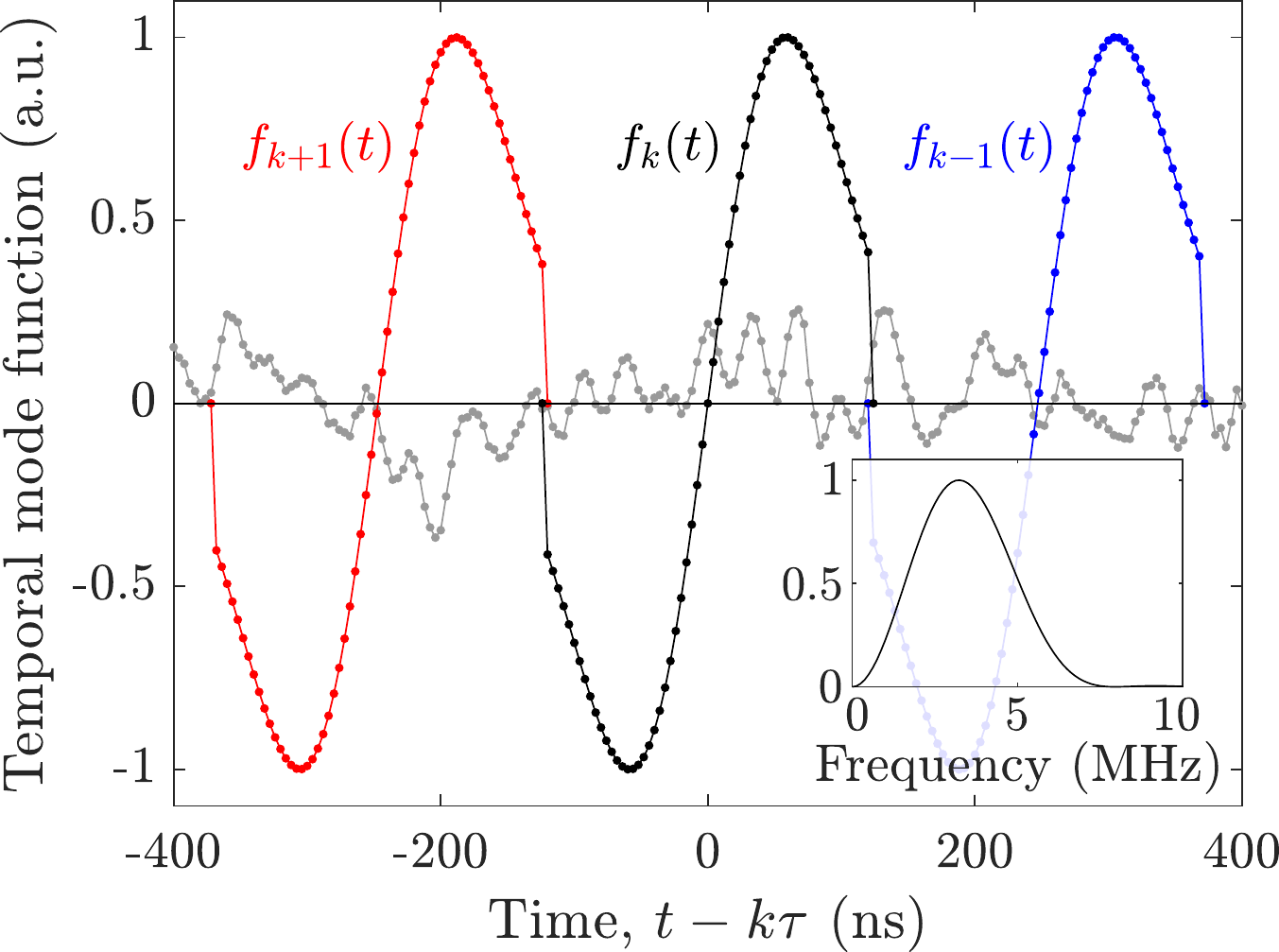}
		\caption{}
	\end{subfigure}
	\begin{subfigure}{0.472\textwidth}
		\includegraphics[width=\textwidth]{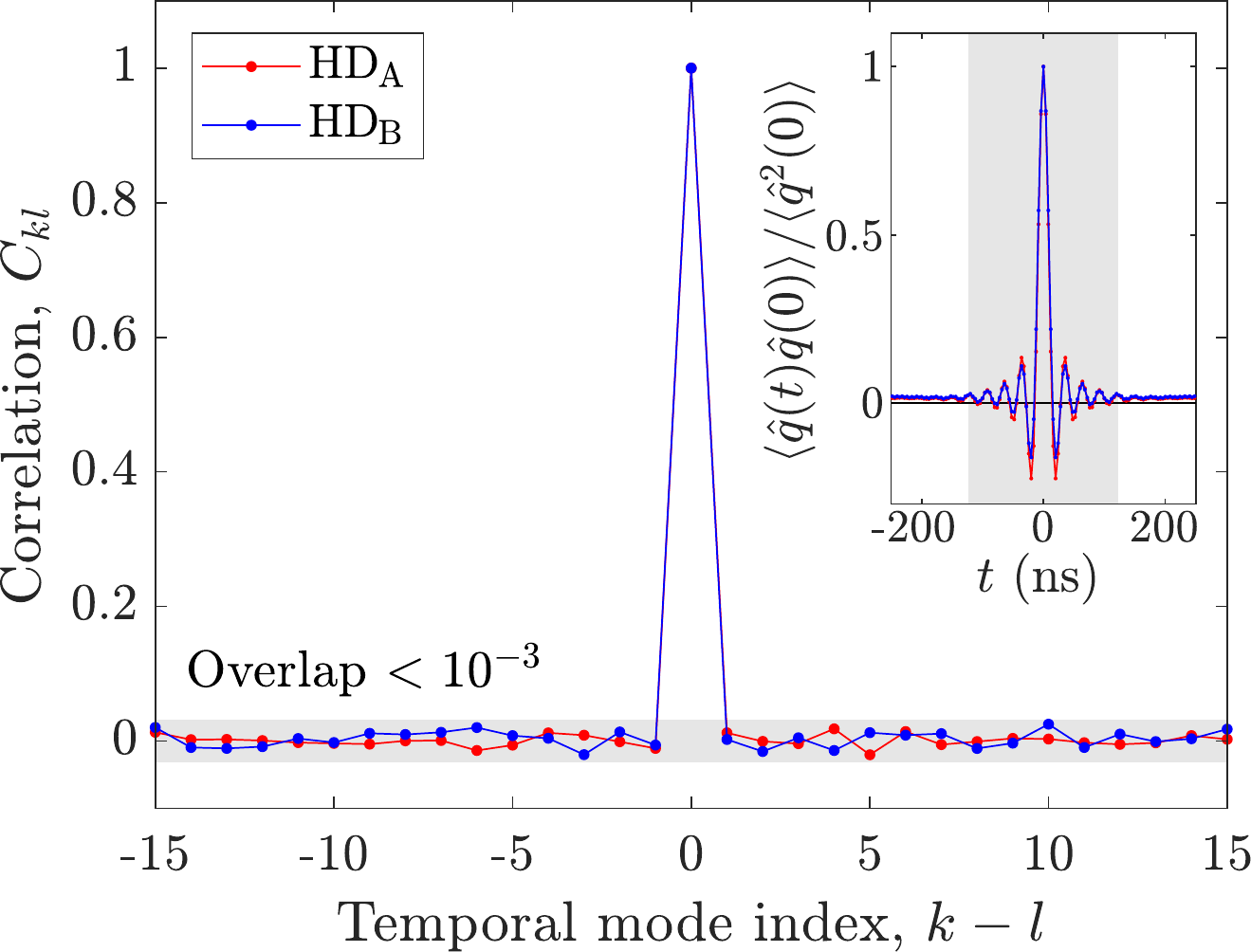}
		\caption{}
	\end{subfigure}
	\caption{(a) Temporal mode function of three neighbouring modes with the form in eq.~(\ref{eq:mode_function}), together with an acquired quadrature time trace (grey). The insert shows the corresponding spectrum of a temporal mode function. (b) Correlations of neighbouring temporal modes with the mode function in (a). Here, the grey area indicates overlap of less than $10^{-3}$. The insert shows the normalized autocorrelation function with the shaded area indicating the time window of a temporal mode.}
	\label{fig:TMF}
\end{figure}

Defined by the short delay length, the temporal mode function is restricted to a temporal window of $\tau=\SI{247}{ns}$ to avoid temporal overlap with neighbouring modes. However, within this window, the shape of the mode function may be optimized to exploit the squeezing spectrum of limited bandwidth and to avoid low frequencies where technical noise dominate. In this work, inspired by {\citeYoshikawa}, we use an uneven temporal mode function given by
\begin{equation}\label{eq:mode_function}
	f_k(t)=\begin{cases}
	    \mathcal{N}(t-k\tau)e^{-\kappa^2(t-k\tau)^2/2}&,\;|t-k\tau|<\frac{\tau}{2}\\
	    0 &,\;\text{otherwise}
	\end{cases}
\end{equation}
where $\mathcal{N}$ is a normalization factor of unit $s^{-1}$, and $\kappa=2\pi\times\SI{2.7}{MHz}$ is optimized to reduce the nullifier variance. Three neighbouring temporal mode functions are shown in Fig.~\ref{fig:TMF}(a) together with an acquired time trace. The mode function is a product of a Gaussian function and a linear term $t-k\tau$: The Gaussian function width defines the mode function bandwidth $\kappa$ which should be within the squeezing source bandwidth, $\gamma_A,\gamma_B$, while the linear term filters out noisy low frequencies. The mode function spectrum is shown in the insert of Fig.~\ref{fig:TMF}(a).

Even though different temporal mode functions do not overlap in time, neighbouring temporal modes may show some overlap due to electronic filtering in the homodyne detectors and electronic noise which can be correlated across multiple temporal modes. To quantify the mode overlap, we measure correlations between different temporal modes of shot noise and the overlap is defined as this correlation squared,
\begin{equation*}
	[\text{Overlap}]=C_{kl}^2=\left(\frac{\braket{\hat{q}_k\hat{q}_l}}{\braket{\hat{q}_k^2}}\right)^2\;.
\end{equation*}
In Fig.~\ref{fig:TMF}(b), correlations between neighbouring modes from a set of $\SI{10000}{}$ quadrature measurements are shown, indicating mode overlap of less than $10^{-3}$. This low overlap is achieved with the uneven mode function where any offset of the acquired data is cancelled, together with little electronic filtering leading to zero autocorrelation outside the temporal mode function window as shown in the insert of Fig.~\ref{fig:TMF}(b).

\section{Results}
Two sets of data are acquired: A small set comprising $\SI{1500}{}$ temporal modes acquired over $\SI{371}{\mu s}$, and a large set of  $\SI{15000}{}$ modes with an acquisition time of $\SI{3.71}{ms}$. Each set includes $\SI{10000}{}$ time traces measured both in the  $\hat{x}$- and $\hat{p}$-basis for building up quadrature statistics to calculate the variances. The sets are acquired with a sampling rate of $\SI{250}{MHz}$ in order to have a large resolution and thus large flexibility in optimizing the delay times.

\begin{figure}
	\centering
	\begin{subfigure}{0.475\textwidth}
		\includegraphics[width=\textwidth]{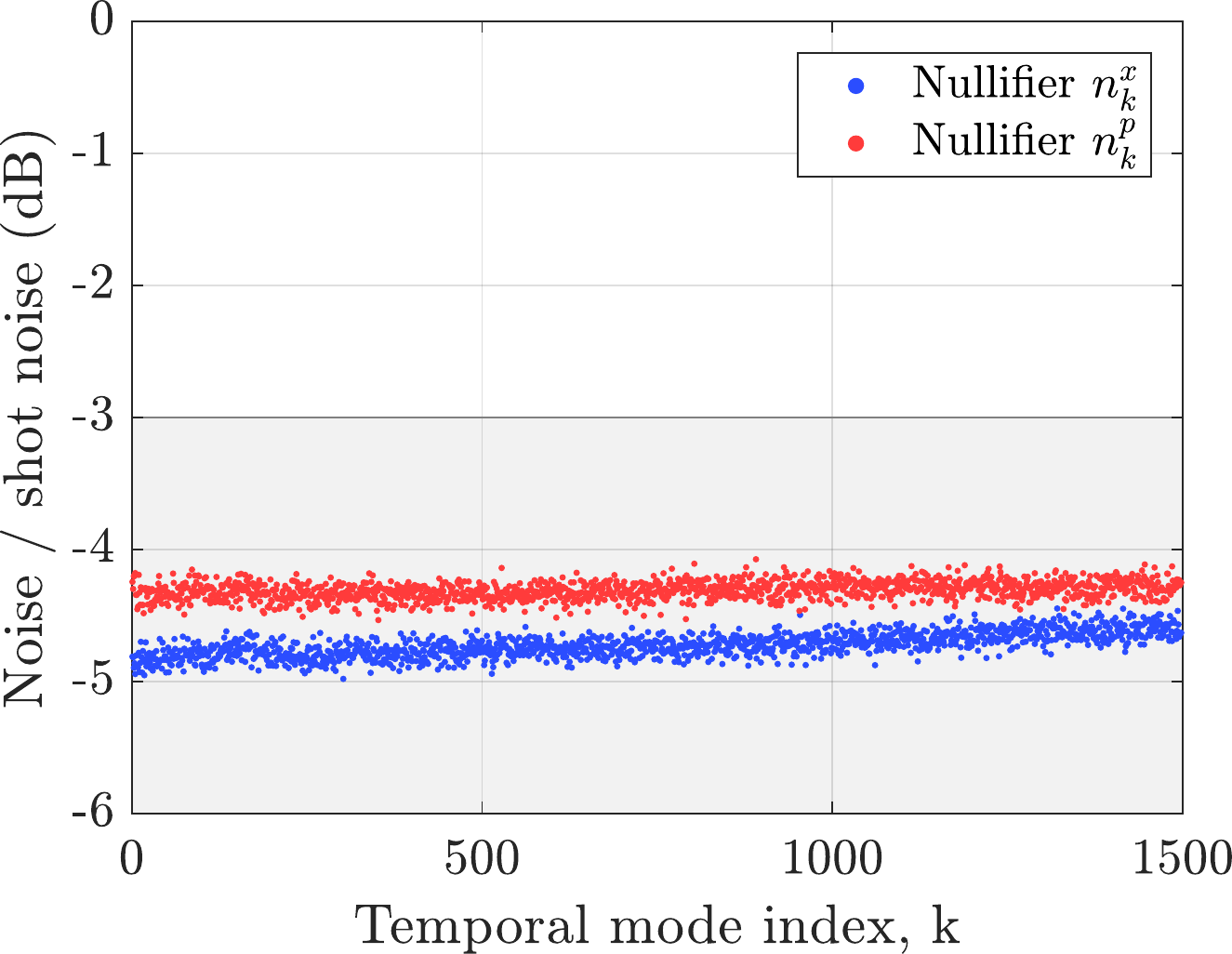}
		\caption{}
	\end{subfigure}
	\begin{subfigure}{0.48\textwidth}
		\includegraphics[width=\textwidth]{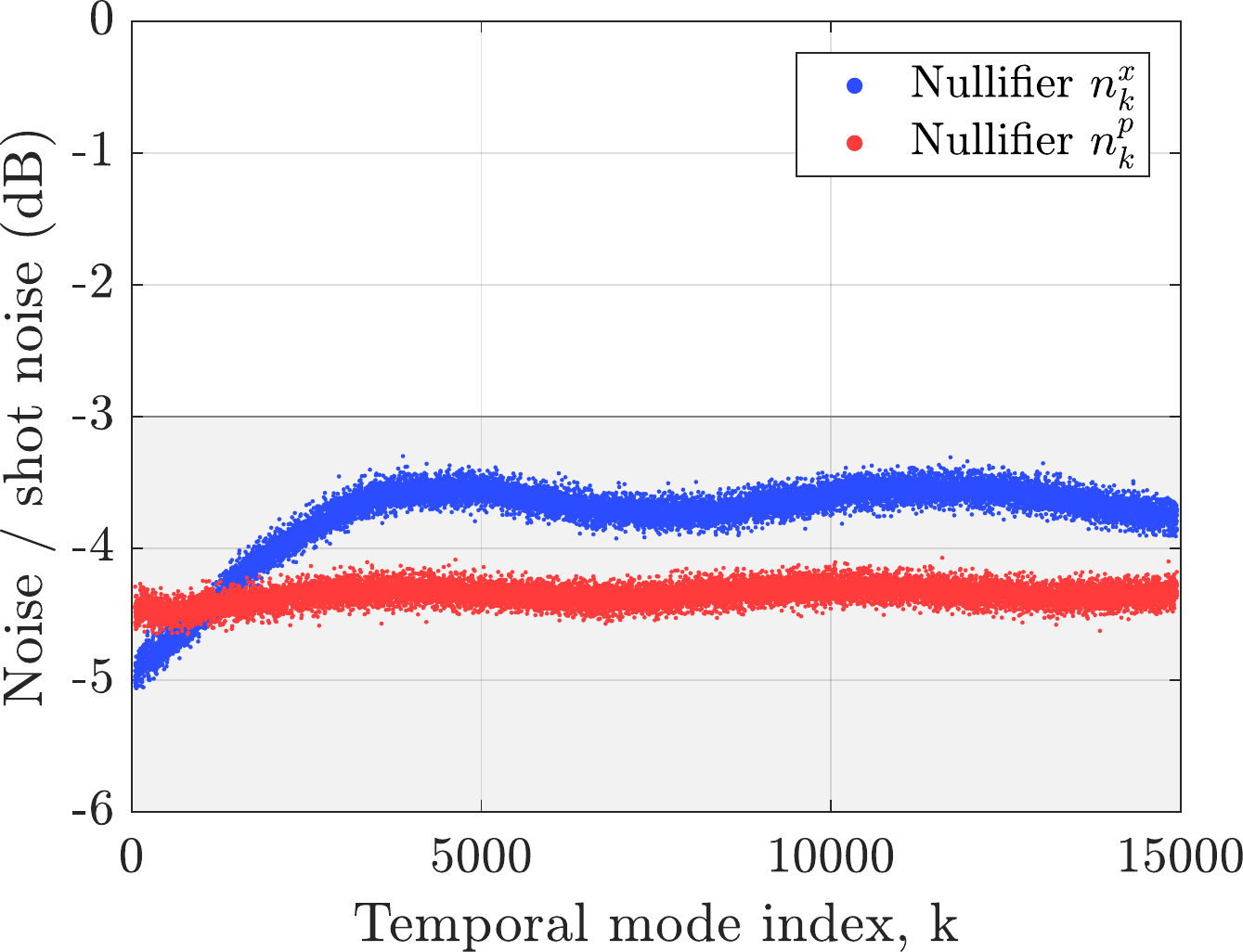}
		\caption{}
	\end{subfigure}
	\caption{Nullifier variance from (a) a short data set of $\SI{371}{\micro s}$ long time traces, and from (b) a long data setup of $\SI{3.71}{ms}$ long time traces. In both figures, the $\SI{-3}{dB}$ separability bound, derived in section \ref{sec:insep} from the van Loock-Furusawa criterion, is marked. With all nullifier variances below this bound, the generated 2D cluster state is completely inseparable.}
	\label{fig:data}
\end{figure}

Using the temporal mode functions described in eq.~(\ref{eq:mode_function}), the $\SI{10000}{}$ quadrature measurements for each temporal mode are extracted from the $\SI{10000}{}$ time traces and normalized to shot noise. Finally, the nullifiers $\hat{n}_k^x$ and $\hat{n}_k^p$ are calculated from the measured quadratures by eq.~(\ref{eq:nullX}-\ref{eq:nullP}) and the nullifier variance is determined. In Fig.~\ref{fig:data}(a) and (b), the resulting nullifier variances are shown for the short and long data set, respectively.

From the short data set, the average variance of $\hat{n}_k^x$ and $\hat{n}_k^p$ is $\SI{-4.7}{dB}$ and $\SI{-4.3}{dB}$ below shot noise, respectively, while the maximum nullifier squeezing measured (an average of 10 neighbouring nullifiers) is $\SI{-4.8}{dB}$ and $\SI{-4.4}{dB}$ respectively. All measured nullifiers show a variance below the $\SI{-3}{dB}$ separability bound derived in section \ref{sec:insep}, and we conclude that the generated 2D cluster state is completely inseparable.
For completeness, we plot in Fig.~S14 the combined variances $\braket{\Delta \hat{X}^2}+\braket{\Delta \hat{P}^2}$ from the van Loock-Furusawa criterion, eq.~\ref{eq:LFc}, for the specific choice of nullifier combinations $\hat{X}$ and $\hat{P}$ used in each of the 127 bipartitions. For all cases, the variances are below the right-hand side of eq.~\ref{eq:LFc}, as indicated by the gray areas, hence explicitly demonstrating the inseparability.

In an attempt to reach a point where the generated cluster state does not violate the $\SI{-3}{dB}$ separability bound due to phase drift when the feedback of cavity and phase locks are kept constant during hold-time, the large data set was acquired. As expected, the nullifier variance increase with time, but even after $\SI{15000}{}$ temporal modes ($\SI{3.71}{ms}$) the phase is stable enough to stay below the separability bound, and we conclude that also the generated $2\times\SI{15000}{}=\SI{30000}{}$ mode (2 spatial modes) 2D cluster state is completely inseparable, while we expect that even larger cluster states may be generated before reaching the separability bound. In the large data set, the average nullifier squeezing of $\hat{n}_k^x$ and $\hat{n}_k^p$ are $\SI{-3.8}{dB}$ and $\SI{-4.4}{dB}$ below shot noise, respectively, while up to (an average of 10 neighbouring nullifiers) $\SI{-5.0}{dB}$ and $\SI{-4.5}{dB}$ of squeezing are measured, respectively.

The periodic variation observed in the nullifier variance in Fig.~\ref{fig:data}(b) is explained by the systematic phase drift from the OPO cavities as discussed in section \ref{sec:eff_and_phase}. We observe a rapid increase of the variance associated with $\hat{n}_k^x$ which may be explained by phase fluctuations of one of the squeezing sources: From eq.~(\ref{eq:quad}) it can be seen that when measuring in the $\hat{x}$-basis, we measure squeezing from the squeezing source in the spatial mode $A$ and anti-squeezing from the the spatial mode $B$ (whereas when measuring in the $\hat{p}$-basis, squeezing and anti-squeezing from the the spatial modes $B$ and $A$ are measured, respectively). When calculating the nullifiers, the anti-squeezing cancels, and we are left with squeezing from one of the two squeezing sources. However, when phases from the squeezing sources drifts, anti-squeezing is mixed into the otherwise squeezed quadrature, and since $\hat{n}_k^x$ only includes squeezing from spatial mode $A$ (and $\hat{n}_k^p$ from spatial mode $B$), we suspect the large relative probe phase fluctuation seen in Fig.~\ref{fig:phase} to be mainly caused by phase drift from the OPO cavity in mode $A$, leading to a rapid increase of $\braket{\Delta\hat{n}_k^{x2}}$ but not $\braket{\Delta\hat{n}_k^{p2}}$. Hence, we expect $\text{OPO}_\text{A}$ to be the dominant source of phase fluctuations that contaminates the measured nullifiers, and not the $\SI{606}{m}$ long fiber delay since we would expect this to affect both $\hat{n}_k^x$ and $\hat{n}_k^p$. Thus, the setup stability may be improved simply by keeping the feedback to cavity locks active at all times. Unfortunately, this was not possible with the current version of the experimental setup, as the cavity lock beams were chopped together with the probe beams in the sample-hold locking scheme described in the Material and Methods section in the beginning of this supplementary material.

\newpage
\noindent\textbf{Table S1:} In the table below, modes of the studied unit cell in the van Loock-Furusawa criterion discussed in the Supplementary Text section \ref{sec:insep} are numbered as indicated in Fig.~\ref{fig:inseparability}. Every bipartition is systematically given an ID between 1 and 127: Consider the 8 bit long binary form of this ID with the least significant bit to the left (e.g. $\text{ID}=3=[1\,1\,0\,0\,0\,0\,0\,0]_\text{binary}$). We then let $\mathcal{S}_1$ include modes with mode number equal to the bit number of bits equal 1 in this binary form of the ID (e.g. $\text{ID}=3\Rightarrow\mathcal{S}_1=\lbrace1,2\rbrace$ and thus $\mathcal{S}_2=\lbrace3,4,5,6,7,8\rbrace$). The table includes Var., the combined variance $\braket{\Delta \hat{X}^2}+\braket{\Delta\hat{P}^2}$; $f$, the right hand side of eq.~(\ref{eq:LFc}); and Sq., the required variance squeezing of each nullifier below shot noise to violate eq.~(\ref{eq:LFc}) with the listed choice of $\hat{X}$ and $\hat{P}$. For clarity, the resulting combined variance $\braket{\Delta \hat{X}^2}+\braket{\Delta \hat{P}^2}$ from acquired data is plotted and compared with $f$ for each bipartition in Fig S14. Note that the squeezing levels listed here are not necessarily the lowest squeezings required to show inseparability for the given bipartition, and for each bipartition a better choice of $\hat{X}$ and $\hat{P}$ may exist which lowers the necessary squeezing.
\begin{longtable}{c  c  c  c c  c  c}
	\hline\hline
	ID & $\mathcal{S}_1$ & $\hat{X}$ & $\hat{P}$ & Var. & $f$ & Sq.\\
	\hline
1 & 1 & $-\hat{n}^x_{k}+\hat{n}^x_{k-1}$ & $-\hat{n}^p_{k}+\hat{n}^p_{k-N}$ & $16e^{-2r}$ & 8 & 3 dB\\ 
2 & 2 & $-\hat{n}^x_{k}+\hat{n}^x_{k-1}$ & $\hat{n}^p_{k}+\hat{n}^p_{k-N}$ & $16e^{-2r}$ & 8 & 3 dB\\ 
3 & 1, 2 & $\hat{n}^x_{k}$ & $\hat{n}^p_{k}$ & $8e^{-2r}$ & 4 & 3 dB\\ 
4 & 3 & $-\hat{n}^x_{k}+\hat{n}^x_{k+1}$ & $\hat{n}^p_{k}+\hat{n}^p_{k-N}$ & $16e^{-2r}$ & 8 & 3 dB\\ 
5 & 1, 3 & $\hat{n}^x_{k}$ & $\hat{n}^p_{k-N}$ & $8e^{-2r}$ & 4 & 3 dB\\ 
6 & 2, 3 & $-\hat{n}^x_{k+1}+\hat{n}^x_{k-1}$ & $\hat{n}^p_{k}+\hat{n}^p_{k-N}$ & $16e^{-2r}$ & 8 & 3 dB\\ 
7 & 1, 2, 3 & $\hat{n}^x_{k}$ & $\hat{n}^p_{k-1}$ & $8e^{-2r}$ & 4 & 3 dB\\ 
8 & 4 & $-\hat{n}^x_{k}+\hat{n}^x_{k+1}$ & $-\hat{n}^p_{k}+\hat{n}^p_{k-N}$ & $16e^{-2r}$ & 8 & 3 dB\\ 
9 & 1, 4 & $-\hat{n}^x_{k+1}+\hat{n}^x_{k-1}$ & $-\hat{n}^p_{k}+\hat{n}^p_{k-N}$ & $16e^{-2r}$ & 8 & 3 dB\\ 
10 & 2, 4 & $\hat{n}^x_{k}$ & $\hat{n}^p_{k-N}$ & $8e^{-2r}$ & 4 & 3 dB\\ 
11 & 1, 2, 4 & $\hat{n}^x_{k}$ & $\hat{n}^p_{k-1}$ & $8e^{-2r}$ & 4 & 3 dB\\ 
12 & 3, 4 & $\hat{n}^x_{k}$ & $\hat{n}^p_{k}$ & $8e^{-2r}$ & 4 & 3 dB\\ 
13 & 1, 3, 4 & $\hat{n}^x_{k}$ & $\hat{n}^p_{k+1}$ & $8e^{-2r}$ & 4 & 3 dB\\ 
14 & 2, 3, 4 & $\hat{n}^x_{k}$ & $\hat{n}^p_{k+1}$ & $8e^{-2r}$ & 4 & 3 dB\\ 
15 & 1, 2, 3, 4 & $\hat{n}^x_{k}$ & $\hat{n}^p_{k+1}$ & $8e^{-2r}$ & 4 & 3 dB\\ 
16 & 5 & $\hat{n}^x_{k}+\hat{n}^x_{k-1}$ & $-\hat{n}^p_{k}+\hat{n}^p_{k+N}$ & $16e^{-2r}$ & 8 & 3 dB\\ 
17 & 1, 5 & $\hat{n}^x_{k}$ & $\hat{n}^p_{k}$ & $8e^{-2r}$ & 4 & 3 dB\\ 
18 & 2, 5 & $\hat{n}^x_{k}$ & $\hat{n}^p_{k}$ & $8e^{-2r}$ & 4 & 3 dB\\ 
19 & 1, 2, 5 & $\hat{n}^x_{k}$ & $\hat{n}^p_{k}$ & $8e^{-2r}$ & 6 & 1.2 dB\\ 
20 & 3, 5 & $-\hat{n}^x_{k}+\hat{n}^x_{k+N}$ & $\hat{n}^p_{k-1}+\hat{n}^p_{k+N}+\hat{n}^p_{k+N+1}+\hat{n}^p_{k+N-1}$ & $24e^{-2r}$ & 12 & 3 dB\\ 
21 & 1, 3, 5 & $\hat{n}^x_{k}$ & $\hat{n}^p_{k-N}$ & $8e^{-2r}$ & 4 & 3 dB\\ 
22 & 2, 3, 5 & $-\hat{n}^x_{k}+\hat{n}^x_{k+N}$ & $-\hat{n}^p_{k}+\hat{n}^p_{k+N}+\hat{n}^p_{k+N+1}+\hat{n}^p_{k+N-1}$ & $24e^{-2r}$ & 12 & 3 dB\\ 
23 & 1, 2, 3, 5 & $\hat{n}^x_{k}$ & $\hat{n}^p_{k}$ & $8e^{-2r}$ & 4 & 3 dB\\ 
24 & 4, 5 & $-\hat{n}^x_{k}+\hat{n}^x_{k+N}$ & $\hat{n}^p_{k-1}+\hat{n}^p_{k+N}+\hat{n}^p_{k+N+1}+\hat{n}^p_{k+N-1}$ & $24e^{-2r}$ & 12 & 3 dB\\ 
25 & 1, 4, 5 & $-\hat{n}^x_{k}+\hat{n}^x_{k+N}$ & $-\hat{n}^p_{k}+\hat{n}^p_{k+N}+\hat{n}^p_{k+N+1}+\hat{n}^p_{k+N-1}$ & $24e^{-2r}$ & 12 & 3 dB\\ 
26 & 2, 4, 5 & $\hat{n}^x_{k}$ & $\hat{n}^p_{k-N}$ & $8e^{-2r}$ & 4 & 3 dB\\ 
27 & 1, 2, 4, 5 & $\hat{n}^x_{k}$ & $\hat{n}^p_{k}$ & $8e^{-2r}$ & 4 & 3 dB\\ 
28 & 3, 4, 5 & $\hat{n}^x_{k}$ & $\hat{n}^p_{k+1}$ & $8e^{-2r}$ & 4 & 3 dB\\ 
29 & 1, 3, 4, 5 & $\hat{n}^x_{k}$ & $\hat{n}^p_{k+1}$ & $8e^{-2r}$ & 4 & 3 dB\\ 
30 & 2, 3, 4, 5 & $\hat{n}^x_{k}$ & $\hat{n}^p_{k+1}$ & $8e^{-2r}$ & 4 & 3 dB\\ 
31 & 1, 2, 3, 4, 5 & $\hat{n}^x_{k}$ & $\hat{n}^p_{k+1}$ & $8e^{-2r}$ & 4 & 3 dB\\ 
32 & 6 & $\hat{n}^x_{k}+\hat{n}^x_{k-1}$ & $\hat{n}^p_{k}+\hat{n}^p_{k+N}$ & $16e^{-2r}$ & 8 & 3 dB\\ 
33 & 1, 6 & $\hat{n}^x_{k}$ & $\hat{n}^p_{k}$ & $8e^{-2r}$ & 4 & 3 dB\\ 
34 & 2, 6 & $\hat{n}^x_{k}$ & $\hat{n}^p_{k}$ & $8e^{-2r}$ & 4 & 3 dB\\ 
35 & 1, 2, 6 & $\hat{n}^x_{k}$ & $\hat{n}^p_{k}$ & $8e^{-2r}$ & 6 & 1.2 dB\\ 
36 & 3, 6 & $\hat{n}^x_{k}+\hat{n}^x_{k+N}$ & $-\hat{n}^p_{k-1}+\hat{n}^p_{k+N}+\hat{n}^p_{k+N+1}+\hat{n}^p_{k+N-1}$ & $24e^{-2r}$ & 12 & 3 dB\\ 
37 & 1, 3, 6 & $\hat{n}^x_{k}$ & $\hat{n}^p_{k-N}$ & $8e^{-2r}$ & 4 & 3 dB\\ 
38 & 2, 3, 6 & $\hat{n}^x_{k}+\hat{n}^x_{k+N}$ & $\hat{n}^p_{k}+\hat{n}^p_{k+N}+\hat{n}^p_{k+N+1}+\hat{n}^p_{k+N-1}$ & $24e^{-2r}$ & 12 & 3 dB\\ 
39 & 1, 2, 3, 6 & $\hat{n}^x_{k}$ & $\hat{n}^p_{k}$ & $8e^{-2r}$ & 4 & 3 dB\\ 
40 & 4, 6 & $\hat{n}^x_{k}+\hat{n}^x_{k+N}$ & $-\hat{n}^p_{k-1}+\hat{n}^p_{k+N}+\hat{n}^p_{k+N+1}+\hat{n}^p_{k+N-1}$ & $24e^{-2r}$ & 12 & 3 dB\\ 
41 & 1, 4, 6 & $\hat{n}^x_{k}+\hat{n}^x_{k+N}$ & $\hat{n}^p_{k}+\hat{n}^p_{k+N}+\hat{n}^p_{k+N+1}+\hat{n}^p_{k+N-1}$ & $24e^{-2r}$ & 12 & 3 dB\\ 
42 & 2, 4, 6 & $\hat{n}^x_{k}$ & $\hat{n}^p_{k-N}$ & $8e^{-2r}$ & 4 & 3 dB\\ 
43 & 1, 2, 4, 6 & $\hat{n}^x_{k}$ & $\hat{n}^p_{k}$ & $8e^{-2r}$ & 4 & 3 dB\\ 
44 & 3, 4, 6 & $\hat{n}^x_{k}$ & $\hat{n}^p_{k+1}$ & $8e^{-2r}$ & 4 & 3 dB\\ 
45 & 1, 3, 4, 6 & $\hat{n}^x_{k}$ & $\hat{n}^p_{k+1}$ & $8e^{-2r}$ & 4 & 3 dB\\ 
46 & 2, 3, 4, 6 & $\hat{n}^x_{k}$ & $\hat{n}^p_{k+1}$ & $8e^{-2r}$ & 4 & 3 dB\\ 
47 & 1, 2, 3, 4, 6 & $\hat{n}^x_{k}$ & $\hat{n}^p_{k+1}$ & $8e^{-2r}$ & 4 & 3 dB\\ 
48 & 5, 6 & $\hat{n}^x_{k}$ & $\hat{n}^p_{k}$ & $8e^{-2r}$ & 4 & 3 dB\\ 
49 & 1, 5, 6 & $\hat{n}^x_{k}$ & $\hat{n}^p_{k}$ & $8e^{-2r}$ & 6 & 1.2 dB\\ 
50 & 2, 5, 6 & $\hat{n}^x_{k}$ & $\hat{n}^p_{k}$ & $8e^{-2r}$ & 6 & 1.2 dB\\ 
51 & 1, 2, 5, 6 & $\hat{n}^x_{k}$ & $\hat{n}^p_{k}$ & $8e^{-2r}$ & 8 & 0 dB\\ 
52 & 3, 5, 6 & $\hat{n}^x_{k}$ & $\hat{n}^p_{k-1}$ & $8e^{-2r}$ & 4 & 3 dB\\ 
53 & 1, 3, 5, 6 & $\hat{n}^x_{k}$ & $\hat{n}^p_{k}$ & $8e^{-2r}$ & 4 & 3 dB\\ 
54 & 2, 3, 5, 6 & $\hat{n}^x_{k}$ & $\hat{n}^p_{k}$ & $8e^{-2r}$ & 4 & 3 dB\\ 
55 & 1, 2, 3, 5, 6 & $\hat{n}^x_{k}$ & $\hat{n}^p_{k}$ & $8e^{-2r}$ & 6 & 1.2 dB\\ 
56 & 4, 5, 6 & $\hat{n}^x_{k}$ & $\hat{n}^p_{k-1}$ & $8e^{-2r}$ & 4 & 3 dB\\ 
57 & 1, 4, 5, 6 & $\hat{n}^x_{k}$ & $\hat{n}^p_{k}$ & $8e^{-2r}$ & 4 & 3 dB\\ 
58 & 2, 4, 5, 6 & $\hat{n}^x_{k}$ & $\hat{n}^p_{k}$ & $8e^{-2r}$ & 4 & 3 dB\\ 
59 & 1, 2, 4, 5, 6 & $\hat{n}^x_{k}$ & $\hat{n}^p_{k}$ & $8e^{-2r}$ & 6 & 1.2 dB\\ 
60 & 3, 4, 5, 6 & $\hat{n}^x_{k}$ & $\hat{n}^p_{k+1}$ & $8e^{-2r}$ & 4 & 3 dB\\ 
61 & 1, 3, 4, 5, 6 & $\hat{n}^x_{k}$ & $\hat{n}^p_{k+1}$ & $8e^{-2r}$ & 4 & 3 dB\\ 
62 & 2, 3, 4, 5, 6 & $\hat{n}^x_{k}$ & $\hat{n}^p_{k+1}$ & $8e^{-2r}$ & 4 & 3 dB\\ 
63 & 1, 2, 3, 4, 5, 6 & $\hat{n}^x_{k}$ & $\hat{n}^p_{k}$ & $8e^{-2r}$ & 4 & 3 dB\\ 
64 & 7 & $\hat{n}^x_{k}+\hat{n}^x_{k+1}$ & $\hat{n}^p_{k}+\hat{n}^p_{k+N}$ & $16e^{-2r}$ & 8 & 3 dB\\ 
65 & 1, 7 & $\hat{n}^x_{k}+\hat{n}^x_{k+N}$ & $-\hat{n}^p_{k+1}+\hat{n}^p_{k+N}+\hat{n}^p_{k+N+1}+\hat{n}^p_{k+N-1}$ & $24e^{-2r}$ & 12 & 3 dB\\ 
66 & 2, 7 & $\hat{n}^x_{k}+\hat{n}^x_{k+N}$ & $-\hat{n}^p_{k+1}+\hat{n}^p_{k+N}+\hat{n}^p_{k+N+1}+\hat{n}^p_{k+N-1}$ & $24e^{-2r}$ & 12 & 3 dB\\ 
67 & 1, 2, 7 & $\hat{n}^x_{k}$ & $\hat{n}^p_{k-1}$ & $8e^{-2r}$ & 4 & 3 dB\\ 
68 & 3, 7 & $\hat{n}^x_{k}$ & $\hat{n}^p_{k}$ & $8e^{-2r}$ & 4 & 3 dB\\ 
69 & 1, 3, 7 & $\hat{n}^x_{k}$ & $\hat{n}^p_{k-N}$ & $8e^{-2r}$ & 4 & 3 dB\\ 
70 & 2, 3, 7 & $\hat{n}^x_{k}+\hat{n}^x_{k+N}$ & $\hat{n}^p_{k}+\hat{n}^p_{k+N}+\hat{n}^p_{k+N+1}+\hat{n}^p_{k+N-1}$ & $24e^{-2r}$ & 12 & 3 dB\\ 
71 & 1, 2, 3, 7 & $\hat{n}^x_{k}$ & $\hat{n}^p_{k-1}$ & $8e^{-2r}$ & 4 & 3 dB\\ 
72 & 4, 7 & $\hat{n}^x_{k}$ & $\hat{n}^p_{k}$ & $8e^{-2r}$ & 4 & 3 dB\\ 
73 & 1, 4, 7 & $\hat{n}^x_{k}+\hat{n}^x_{k+N}$ & $\hat{n}^p_{k}+\hat{n}^p_{k+N}+\hat{n}^p_{k+N+1}+\hat{n}^p_{k+N-1}$ & $24e^{-2r}$ & 12 & 3 dB\\ 
74 & 2, 4, 7 & $\hat{n}^x_{k}$ & $\hat{n}^p_{k-N}$ & $8e^{-2r}$ & 4 & 3 dB\\ 
75 & 1, 2, 4, 7 & $\hat{n}^x_{k}$ & $\hat{n}^p_{k-1}$ & $8e^{-2r}$ & 4 & 3 dB\\ 
76 & 3, 4, 7 & $\hat{n}^x_{k}$ & $\hat{n}^p_{k}$ & $8e^{-2r}$ & 6 & 1.2 dB\\ 
77 & 1, 3, 4, 7 & $\hat{n}^x_{k}$ & $\hat{n}^p_{k}$ & $8e^{-2r}$ & 4 & 3 dB\\ 
78 & 2, 3, 4, 7 & $\hat{n}^x_{k}$ & $\hat{n}^p_{k}$ & $8e^{-2r}$ & 4 & 3 dB\\ 
79 & 1, 2, 3, 4, 7 & $\hat{n}^x_{k}$ & $\hat{n}^p_{k-1}$ & $8e^{-2r}$ & 4 & 3 dB\\ 
80 & 5, 7 & $\hat{n}^x_{k}$ & $\hat{n}^p_{k+N}$ & $8e^{-2r}$ & 4 & 3 dB\\ 
81 & 1, 5, 7 & $\hat{n}^x_{k}$ & $\hat{n}^p_{k+N}$ & $8e^{-2r}$ & 4 & 3 dB\\ 
82 & 2, 5, 7 & $\hat{n}^x_{k}$ & $\hat{n}^p_{k+N}$ & $8e^{-2r}$ & 4 & 3 dB\\ 
83 & 1, 2, 5, 7 & $\hat{n}^x_{k}$ & $\hat{n}^p_{k}$ & $8e^{-2r}$ & 4 & 3 dB\\ 
84 & 3, 5, 7 & $\hat{n}^x_{k}$ & $\hat{n}^p_{k+N}$ & $8e^{-2r}$ & 4 & 3 dB\\ 
85 & 1, 3, 5, 7 & $\hat{n}^x_{k}$ & $\hat{n}^p_{k+N}$ & $8e^{-2r}$ & 4 & 3 dB\\ 
86 & 2, 3, 5, 7 & $\hat{n}^x_{k}$ & $\hat{n}^p_{k+N}$ & $8e^{-2r}$ & 4 & 3 dB\\ 
87 & 1, 2, 3, 5, 7 & $\hat{n}^x_{k}$ & $\hat{n}^p_{k+N}$ & $8e^{-2r}$ & 4 & 3 dB\\ 
88 & 4, 5, 7 & $\hat{n}^x_{k}$ & $\hat{n}^p_{k+N}$ & $8e^{-2r}$ & 4 & 3 dB\\ 
89 & 1, 4, 5, 7 & $\hat{n}^x_{k}$ & $\hat{n}^p_{k+N}$ & $8e^{-2r}$ & 4 & 3 dB\\ 
90 & 2, 4, 5, 7 & $\hat{n}^x_{k}$ & $\hat{n}^p_{k+N}$ & $8e^{-2r}$ & 4 & 3 dB\\ 
91 & 1, 2, 4, 5, 7 & $\hat{n}^x_{k}$ & $\hat{n}^p_{k+N}$ & $8e^{-2r}$ & 4 & 3 dB\\ 
92 & 3, 4, 5, 7 & $\hat{n}^x_{k}$ & $\hat{n}^p_{k}$ & $8e^{-2r}$ & 4 & 3 dB\\ 
93 & 1, 3, 4, 5, 7 & $\hat{n}^x_{k}$ & $\hat{n}^p_{k+N}$ & $8e^{-2r}$ & 4 & 3 dB\\ 
94 & 2, 3, 4, 5, 7 & $\hat{n}^x_{k}$ & $\hat{n}^p_{k+N}$ & $8e^{-2r}$ & 4 & 3 dB\\ 
95 & 1, 2, 3, 4, 5, 7 & $\hat{n}^x_{k}$ & $\hat{n}^p_{k+N}$ & $8e^{-2r}$ & 4 & 3 dB\\ 
96 & 6, 7 & $-\hat{n}^x_{k+1}+\hat{n}^x_{k-1}$ & $\hat{n}^p_{k}+\hat{n}^p_{k+N}$ & $16e^{-2r}$ & 8 & 3 dB\\ 
97 & 1, 6, 7 & $-\hat{n}^x_{k}+\hat{n}^x_{k-N}$ & $\hat{n}^p_{k}-\hat{n}^p_{k-N}+\hat{n}^p_{k-N+1}+\hat{n}^p_{k-N-1}$ & $24e^{-2r}$ & 12 & 3 dB\\ 
98 & 2, 6, 7 & $\hat{n}^x_{k}+\hat{n}^x_{k-N}$ & $-\hat{n}^p_{k}-\hat{n}^p_{k-N}+\hat{n}^p_{k-N+1}+\hat{n}^p_{k-N-1}$ & $24e^{-2r}$ & 12 & 3 dB\\ 
99 & 1, 2, 6, 7 & $\hat{n}^x_{k}$ & $\hat{n}^p_{k}$ & $8e^{-2r}$ & 4 & 3 dB\\ 
100 & 3, 6, 7 & $\hat{n}^x_{k}+\hat{n}^x_{k-N}$ & $-\hat{n}^p_{k}-\hat{n}^p_{k-N}+\hat{n}^p_{k-N+1}+\hat{n}^p_{k-N-1}$ & $24e^{-2r}$ & 12 & 3 dB\\ 
101 & 1, 3, 6, 7 & $\hat{n}^x_{k}$ & $\hat{n}^p_{k-N}$ & $8e^{-2r}$ & 4 & 3 dB\\ 
102 & 2, 3, 6, 7 & $-\hat{n}^x_{k+1}+\hat{n}^x_{k-1}$ & $-\hat{n}^p_{k+N}+\hat{n}^p_{k-N}$ & $16e^{-2r}$ & 8 & 3 dB\\ 
103 & 1, 2, 3, 6, 7 & $-\hat{n}^x_{k}+\hat{n}^x_{k-N}$ & $\hat{n}^p_{k}-\hat{n}^p_{k-N}+\hat{n}^p_{k-N+1}+\hat{n}^p_{k-N-1}$ & $24e^{-2r}$ & 12 & 3 dB\\ 
104 & 4, 6, 7 & $-\hat{n}^x_{k}+\hat{n}^x_{k-N}$ & $\hat{n}^p_{k}-\hat{n}^p_{k-N}+\hat{n}^p_{k-N+1}+\hat{n}^p_{k-N-1}$ & $24e^{-2r}$ & 12 & 3 dB\\ 
105 & 1, 4, 6, 7 & $-\hat{n}^x_{k+1}+\hat{n}^x_{k-1}$ & $\hat{n}^p_{k+N}+\hat{n}^p_{k-N}$ & $16e^{-2r}$ & 8 & 3 dB\\ 
106 & 2, 4, 6, 7 & $\hat{n}^x_{k}$ & $\hat{n}^p_{k-N}$ & $8e^{-2r}$ & 4 & 3 dB\\ 
107 & 1, 2, 4, 6, 7 & $\hat{n}^x_{k}+\hat{n}^x_{k-N}$ & $-\hat{n}^p_{k}-\hat{n}^p_{k-N}+\hat{n}^p_{k-N+1}+\hat{n}^p_{k-N-1}$ & $24e^{-2r}$ & 12 & 3 dB\\ 
108 & 3, 4, 6, 7 & $\hat{n}^x_{k}$ & $\hat{n}^p_{k}$ & $8e^{-2r}$ & 4 & 3 dB\\ 
109 & 1, 3, 4, 6, 7 & $\hat{n}^x_{k}+\hat{n}^x_{k-N}$ & $-\hat{n}^p_{k}-\hat{n}^p_{k-N}+\hat{n}^p_{k-N+1}+\hat{n}^p_{k-N-1}$ & $24e^{-2r}$ & 12 & 3 dB\\ 
110 & 2, 3, 4, 6, 7 & $-\hat{n}^x_{k}+\hat{n}^x_{k-N}$ & $\hat{n}^p_{k}-\hat{n}^p_{k-N}+\hat{n}^p_{k-N+1}+\hat{n}^p_{k-N-1}$ & $24e^{-2r}$ & 12 & 3 dB\\ 
111 & 1, 2, 3, 4, 6, 7 & $-\hat{n}^x_{k+1}+\hat{n}^x_{k-1}$ & $-\hat{n}^p_{k}+\hat{n}^p_{k+N}$ & $16e^{-2r}$ & 8 & 3 dB\\ 
112 & 5, 6, 7 & $\hat{n}^x_{k}$ & $\hat{n}^p_{k-1}$ & $8e^{-2r}$ & 4 & 3 dB\\ 
113 & 1, 5, 6, 7 & $\hat{n}^x_{k}$ & $\hat{n}^p_{k}$ & $8e^{-2r}$ & 4 & 3 dB\\ 
114 & 2, 5, 6, 7 & $\hat{n}^x_{k}$ & $\hat{n}^p_{k}$ & $8e^{-2r}$ & 4 & 3 dB\\ 
115 & 1, 2, 5, 6, 7 & $\hat{n}^x_{k}$ & $\hat{n}^p_{k}$ & $8e^{-2r}$ & 6 & 1.2 dB\\ 
116 & 3, 5, 6, 7 & $\hat{n}^x_{k}$ & $\hat{n}^p_{k-1}$ & $8e^{-2r}$ & 4 & 3 dB\\ 
117 & 1, 3, 5, 6, 7 & $\hat{n}^x_{k}$ & $\hat{n}^p_{k-N}$ & $8e^{-2r}$ & 4 & 3 dB\\ 
118 & 2, 3, 5, 6, 7 & $-\hat{n}^x_{k}+\hat{n}^x_{k+N}$ & $-\hat{n}^p_{k}+\hat{n}^p_{k+N}+\hat{n}^p_{k+N+1}+\hat{n}^p_{k+N-1}$ & $24e^{-2r}$ & 12 & 3 dB\\ 
119 & 1, 2, 3, 5, 6, 7 & $\hat{n}^x_{k}$ & $\hat{n}^p_{k}$ & $8e^{-2r}$ & 4 & 3 dB\\ 
120 & 4, 5, 6, 7 & $\hat{n}^x_{k}$ & $\hat{n}^p_{k-1}$ & $8e^{-2r}$ & 4 & 3 dB\\ 
121 & 1, 4, 5, 6, 7 & $-\hat{n}^x_{k}+\hat{n}^x_{k+N}$ & $-\hat{n}^p_{k}+\hat{n}^p_{k+N}+\hat{n}^p_{k+N+1}+\hat{n}^p_{k+N-1}$ & $24e^{-2r}$ & 12 & 3 dB\\ 
122 & 2, 4, 5, 6, 7 & $\hat{n}^x_{k}$ & $\hat{n}^p_{k-N}$ & $8e^{-2r}$ & 4 & 3 dB\\ 
123 & 1, 2, 4, 5, 6, 7 & $\hat{n}^x_{k}$ & $\hat{n}^p_{k}$ & $8e^{-2r}$ & 4 & 3 dB\\ 
124 & 3, 4, 5, 6, 7 & $\hat{n}^x_{k}$ & $\hat{n}^p_{k-1}$ & $8e^{-2r}$ & 4 & 3 dB\\ 
125 & 1, 3, 4, 5, 6, 7 & $-\hat{n}^x_{k}+\hat{n}^x_{k+N}$ & $\hat{n}^p_{k+1}+\hat{n}^p_{k+N}+\hat{n}^p_{k+N+1}+\hat{n}^p_{k+N-1}$ & $24e^{-2r}$ & 12 & 3 dB\\ 
126 & 2, 3, 4, 5, 6, 7 & $-\hat{n}^x_{k}+\hat{n}^x_{k+N}$ & $\hat{n}^p_{k+1}+\hat{n}^p_{k+N}+\hat{n}^p_{k+N+1}+\hat{n}^p_{k+N-1}$ & $24e^{-2r}$ & 12 & 3 dB\\ 
127 & 1, 2, 3, 4, 5, 6, 7 & $\hat{n}^x_{k}+\hat{n}^x_{k+1}$ & $-\hat{n}^p_{k}+\hat{n}^p_{k+N}$ & $16e^{-2r}$ & 8 & 3 dB\\ 
	\hline\hline
\end{longtable}

\newpage
\noindent\textbf{Fig. S14:} Plot of the van Loock-Furusawa separability criterion's left-hand-side, $\braket{\Delta \hat{X}^2}+\braket{\Delta \hat{P}^2}$, for each bipartition and chosen nullifier combination listed in table S1 using acquired data. Here, the gray area marks values less than the van Loock-Furusawa criterion right-hand-side, $f$, listed in table S1, and corresponds to the given bipartition being inseparable.
\begin{center}
    \includegraphics[width=0.95\textwidth]{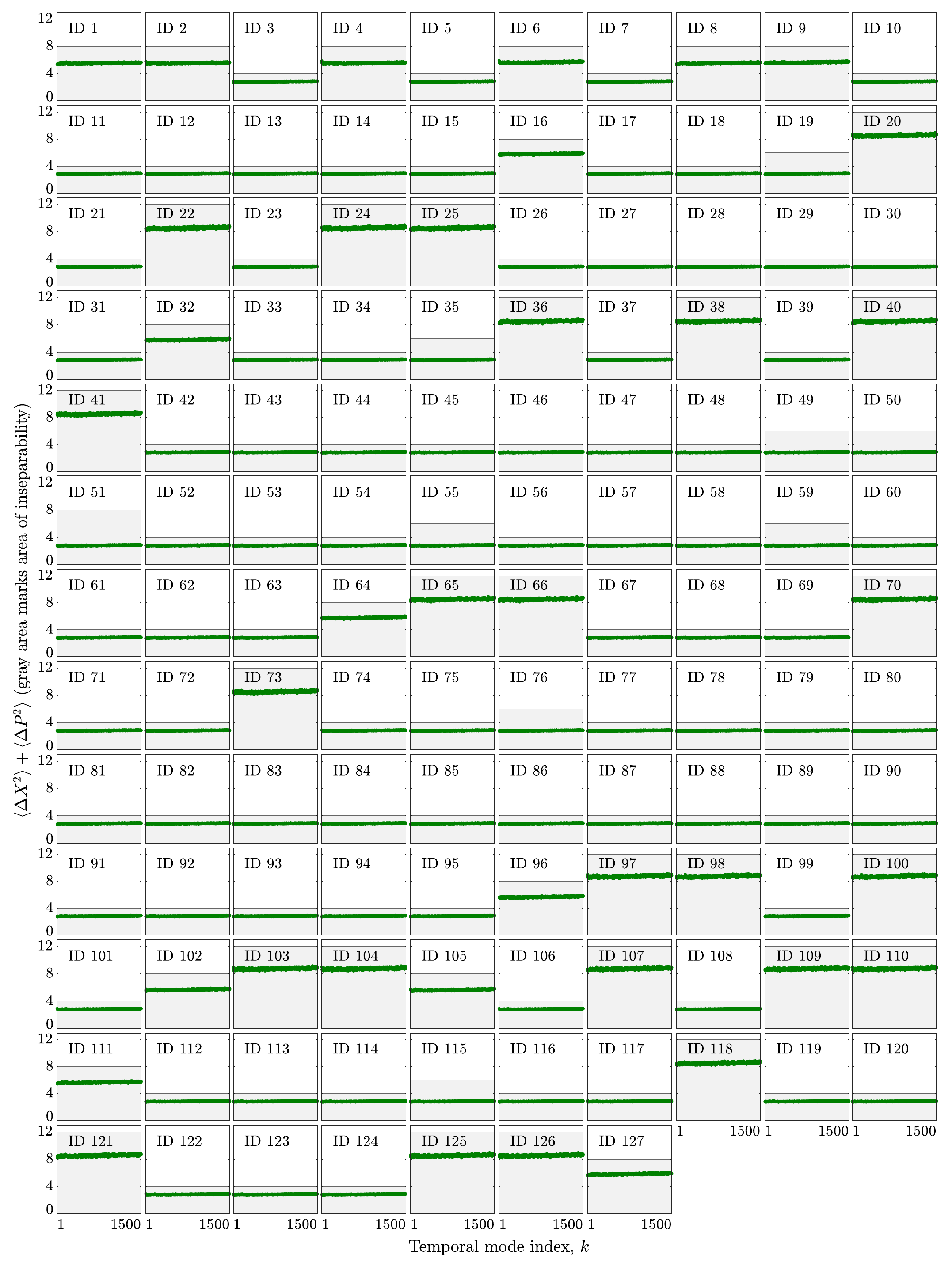}
\end{center}